\documentclass[preprint]{ptephy_v2}
\preprintnumber{YITP-26-84} 

\pdfoutput=1 

\usepackage[T1]{fontenc}
\usepackage[utf8]{inputenc}
\usepackage{graphicx}
\usepackage{dcolumn}
\usepackage{bm}
\usepackage{mathtools}
\usepackage{slashed}
\usepackage[usenames,dvipsnames,svgnames,table,x11names]{xcolor}
\usepackage{extarrows}
\usepackage{easybmat}
\usepackage{tensor}
\usepackage{nicematrix}
\usepackage{adjustbox}

\definecolor{light_blue}{rgb}{0.15, 0.35, 0.95}
\definecolor{kit_green}{rgb}{0, 
0.58823 
, 0.50980 
}

\usepackage{multirow}
\usepackage{booktabs} 
\usepackage{siunitx} 
\usepackage{tabularx} 

\newcommand\subsetsim{\mathrel{%
  \ooalign{\raise0.2ex\hbox{$\subset$}\cr\hidewidth\raise-0.8ex\hbox{\scalebox{0.9}{$\sim$}}\hidewidth\cr}}}
\newcommand\supsetsim{\mathrel{%
  \ooalign{\raise0.2ex\hbox{$\supset$}\cr\hidewidth\raise-0.8ex\hbox{\scalebox{0.9}{$\sim$}}\hidewidth\cr}}}

\newcommand{\olr}[1]{\overset{\text{\scriptsize$\leftrightarrow$}}{#1}}
\newcommand{\slasholr}[1]{{\,\slashed{#1}\mathllap{\olr{\phantom{#1}}}\,}}
\newcommand{\ol}[1]{\overset{\text{\scriptsize$\leftarrow$}}{#1}}
\newcommand{\slashol}[1]{{\,\slashed{#1}\mathllap{\ol{\phantom{#1}}}\,}}
\newcommand{\oPsi}{\overline{\Psi}}
\newcommand{\opsi}{\overline{\psi}}
\newcommand{\Tr}{\text{Tr}}
\newcommand{\oh}{\overline{h}}
\newcommand{\ov}[1]{\overline{#1}}

\newcommand{\mA}{\mathcal{A}}
\newcommand{\mB}{\mathcal{B}}
\renewcommand{\det}{\text{det}}
\newcommand{\wGamma}{\widetilde{\Gamma}}

\usepackage{makecell}
\setcellgapes{6pt} 
\makegapedcells    

\usepackage[bookmarksopen,colorlinks=true,linkcolor=blue,citecolor=blue,urlcolor=blue]{hyperref}

\allowdisplaybreaks[1]

\begin{document}

\title{Kaluza--Klein tower thresholds and scheme dependence of the species scale}

\author{Yuri Michinobu}
\affil{Center for Gravitational Physics and Quantum Information, Yukawa Institute for Theoretical Physics, Kyoto University, Kitashirakawa Oiwakecho, Sakyo-ku, Kyoto 606-8502, Japan \email{yuri.michinobu@yukawa.kyoto-u.ac.jp}}


\begin{abstract}
We revisit the species scale in quantum gravity from the viewpoint of effective field theory (EFT). Two common parametric characterizations relate it to the breakdown of perturbative gravity and to the suppression scale of higher-derivative gravitational operators. We examine their correspondence through local threshold corrections from infinite Kaluza--Klein (KK) towers. In four dimensions, we compute the four-derivative one-loop action of minimally coupled scalar and Dirac towers with KK spectra in graviton, radion, and photon backgrounds. Feynman diagrams determine the single-mode coefficients, while analytic summation with full threshold functions tests the logarithmic coefficient inferred from renormalization-group evolution. The leading contribution scales with the effective number of KK levels, reproducing the standard species-scale relation under the stated matching assumptions, but its coefficient depends on the cutoff profile and belongs to EFT matching data. In contrast, the subleading curvature-squared thresholds contain logarithms of the KK mass gap whose spin-dependent coefficients remain stable under changes of the cutoff profile within the class studied here.
\end{abstract}

\maketitle
\flushbottom

\section{Introduction}\label{sec:intro}

A large number of light species lowers the scale at which perturbative gravity is expected to break down. In four dimensions, the standard parametric relation is
\begin{equation}
    \Lambda_{\mathrm{sp}}\sim
    \frac{M_{\mathrm{pl},4}}{\sqrt{N}},
\end{equation}
up to order-one factors and possible logarithmic corrections
\cite{Dvali:2007hz,Dvali:2007wp,Dvali:2009ks,Dvali:2010vm,
Veneziano:2001ah,Han:2004wt,Dvali:2012uq,Aydemir:2012nz,
Calmet:2017omb,Castellano:2022bvr,Long:2021jlv}.
For a tower of states, $N$ depends on the scale at which it is evaluated, so this relation must be imposed self-consistently, as discussed in~\cite{Castellano:2022bvr,Castellano:2023jjt} and explicit one-loop tower calculations
\cite{Cribiori:2023ffn,Aoufia:2024awo}. A recent effective field theory (EFT) treatment of this self-consistent tower counting was given in
\cite{ValeixoBento:2025bmv}. Our analysis is complementary: rather than deriving the species scale from the self-consistent counting condition itself, we ask how the corresponding Kaluza--Klein (KK) tower contribution is encoded in local higher-derivative matching data, and which part of that data is stable under changes of the cutoff profile.

A second characterization relates the species scale to the suppression scale of higher-derivative terms in the gravitational effective action
\cite{vandeHeisteeg:2022btw,Cribiori:2022nke,Castellano:2023aum}.
This perspective is useful in compactifications with moduli-dependent towers and has been developed using geometric, moduli-space, black-hole, and EFT formulations
\cite{vandeHeisteeg:2023ubh,vandeHeisteeg:2023dlw,
Calderon-Infante:2023uhz,Castellano:2023stg,
Cribiori:2023sch,Bedroya:2024ubj,Bedroya:2024uva,
Castellano:2024bna,Castellano:2025ljk,ValeixoBento:2025bmv}.
More recently, Laplace-type equations obeyed by protected higher-derivative Wilson coefficients have been reinterpreted as constraints on moduli-dependent species-scale data and species hull vectors~\cite{Aoufia:2025ppe}. Explicit heat-kernel, free-energy, and amplitude calculations provide complementary probes~\cite{Cribiori:2023ffn,Aoufia:2024awo,Basile:2024dqq,Herraez:2024kux}. An amplitude-based characterization of the gravitational cutoff, based on unitarity and analyticity of graviton scattering, has also been proposed~\cite{Caron-Huot:2024lbf}; this should be contrasted with the local matching problem studied below.

The perturbative-breakdown and higher-derivative characterizations of the species scale are both intended as parametric statements. Since higher-derivative coefficients are local EFT matching coefficients, their finite parts are generally sensitive to regulator and subtraction schemes. Such scheme dependence is already expected from standard EFT reasoning and is consistent with the parametric status of both characterizations. The propagator-based argument also has known limitations: resumming a one-loop self-energy alone does not establish a physical gravitational cutoff~\cite{Caron-Huot:2024lbf}. These considerations motivate a more specific question: how does the choice of cutoff profile affect the four-derivative coefficients generated by an infinite KK tower, and which parts of their large-$N$ expansion remain unchanged?

In this paper, we address this question for an explicitly regulated infinite KK sum. We show that the tower-induced local threshold correction separates into a regulator-dependent leading term and a logarithmic contribution whose coefficient is stable within the class of proper-time cutoff profiles considered below. The analysis has two parts: an analytic treatment of the regulated tower sum determines its cutoff-profile dependence and the conditions on the regulator, while explicit Feynman-diagram calculations determine the single-mode coefficients for scalar and fermion fields. Combining the two gives the four-derivative threshold coefficients of the KK towers.

A local loop contribution depends on the regulator and subtraction prescription and does not, by itself, determine the full Wilson coefficient. At a specified renormalization scale, it must be combined with the corresponding counterterm and renormalized matching coefficient
\cite{Manohar:2018aog,Burgess:2020tbq}. Physical amplitudes or correlation functions of consistently matched operators are independent of this bookkeeping only after the Wilson coefficients and loop matrix elements are treated consistently
\cite{Donoghue:1994dn,Donoghue:2022eay}; field redefinitions may also change the displayed operator basis without changing observables
\cite{Criado:2018sdb}.
The tower calculation supplies the local threshold data for this matching, with their cutoff-profile dependence made explicit. This allows one to distinguish the generic tower enhancement from the numerical normalization entering higher-derivative estimates of the species scale.

We study these issues in four-dimensional EFTs of minimally coupled scalar and Dirac towers with the KK spectrum $m_{{\rm KK},n}=n\,m_{\rm KK}, n\geq 1$. Their masses, charges, and radion dependence are chosen as in an $S^1$ compactification. The calculation integrates out the nonzero massive matter modes at one loop, treating the graviton, radion, and photon as background fields. The massive KK modes of the higher-dimensional metric are outside the field content of these models. For the fermion tower, Pauli couplings to the photon are set to zero as part of the model definition, whose relation to circle compactification is explained in Appendix~\ref{app:cc}.

As the KK gap decreases, these models capture the light-tower behavior characteristic of the decompactification regime, in line with the Swampland Distance Conjecture and related refinements
\cite{Ooguri:2006in,Grimm:2018ohb,Heidenreich:2017sim,Heidenreich:2018kpg,
Montero:2022prj,Etheredge:2023odp,Etheredge:2024tok,Blumenhagen:2023xmk}. A constant radion background $\phi_0$ fixes the KK gap, and a provisional ultraviolet scale $\Lambda_{\mathrm{sp},0}$ defines
\begin{equation}
    N(\phi_0)=
    \frac{\Lambda_{\mathrm{sp},0}}{m_{\rm KK}(\phi_0)}.
\end{equation}
Here $N(\phi_0)$ denotes the effective number of levels of the chosen matter tower below the provisional cutoff scale; it is not a count of additional spin, polarization, charge, or internal degeneracy factors. We extract four-derivative terms involving the graviton, radion, and KK photon. In the final four-dimensional formulas, terms containing an undifferentiated radion are omitted, as in the body of the paper.

For the scalar tower, the one-loop effective action can be computed directly with heat kernel methods~\cite{Vassilevich:2003xt,Cribiori:2023ffn,Aoufia:2024awo,Gilkey:1995mj,Avramidi:2000bm,Avramidi:1990je,Avramidi:1990ug}. For the fermion tower, however, a direct heat-kernel treatment is less straightforward because the KK mass depends on the radion and naive doubling of the Dirac operator introduces a possible multiplicative anomaly~\cite{DeBerredo-Peixoto:2001wkv,Kontsevich:1994xe,Goncalves:2009sk,Elizalde:1997nd,Elizalde:1998vd,Elizalde:1999zy,Cognola:1999xv}. We therefore use a factorized diagrammatic prescription: Pauli--Villars regularization defines the finite part of each single-mode diagram, while a proper-time cutoff regulates the infinite KK sum. Earlier work documents regulator and ultraviolet sensitivity in compactified theories
\cite{Branchina:2023rgi,Contino:2001gz}, but the precise factorized prescription adopted here is a choice of the present calculation.
This prescription allows the scalar and fermion towers to be treated in parallel. In the scalar case, one can explicitly verify that the result obtained with this prescription agrees with the heat-kernel result. The purpose of the prescription is to expose the tower dependence and the cutoff-profile dependence of the local terms.

The central technical result is the large-$N$ behavior of the regulated tower sum. For a common proper-time cutoff function $f_c(u)$, held fixed as $N$ varies and satisfying the convergence conditions discussed below, the tower sum has the form
\begin{equation}
    \label{eq:tower_cutoff_structure}
    \mathcal{N}_{a+b} [f_c] = \frac{1}{2} g_{a+b} \,\widetilde{f}_c(-1/2) N(\phi_0) + \delta_{a+b,0} \,\mathrm{Res}_{s=0} \widetilde{f}_c \cdot \log N(\phi_0) + \mathcal{O}(N^0(\phi_0)).
\end{equation}
Here $\widetilde{f}_c(s)$ is the Mellin transform of $f_c(u)$, while $a$, $b$, and $g_{a+b}$ are defined in \autoref{sec:factorized_scheme}. The coefficient of the leading local term depends on the cutoff profile through the value of the Mellin  transform at $s=-1/2$, $\widetilde{f}_c(-1/2)$, the same factor for all values of $a+b$. The tower sum therefore gives a generic parametric $N$ enhancement with a common cutoff-dependent normalization. By contrast, in the curvature-squared sector $(a+b=0)$ the coefficient of $\log N(\phi_0)$ is invariant within this cutoff class because ${\rm Res}_{s=0}\widetilde{f}_c(s)=-1$.

The logarithmic term is governed by the same single-mode logarithmic divergence that enters ordinary one-loop renormalization. As explained in \autoref{sec:universality}, the threshold logarithms can be understood by including each KK mode's contribution to the beta function above its mass threshold and integrating the RG equation. The result computed here is a local threshold correction; a nonanalytic momentum-space logarithm such as $\log(-p^2)$ belongs to a finite-momentum calculation.

The explicit four-dimensional scalar and fermion results exhibit this same separation. Both contain cutoff-profile dependent local contributions proportional to $N(\phi_0)$ and spin-dependent logarithmic curvature-squared terms whose $\log N(\phi_0)$ coefficients are fixed within the analyzed cutoff class. After a field redefinition, the radion and graviton sectors can be displayed in a basis containing $(\nabla\phi)^4$, giving the corresponding local matching coefficients in that basis.

The leading enhancement is parametrically consistent with the standard species scaling, while the logarithmic terms retain information about the KK mass gap and the spin of the tower. Our results identify this logarithmic information independently of the cutoff-dependent normalization of the leading term, allowing scalar and fermion thresholds to be compared beyond their leading species enhancement.

The paper is organized as follows. \autoref{sec:species} reviews the perturbative and higher-derivative characterizations of the species scale. \autoref{sec:universality} analyzes the proper-time cutoff dependence and the role of local matching coefficients. \autoref{sec:one_loop_scalar} computes one-loop corrections from the scalar tower with heat kernel methods and derives the parametric self-consistency condition. \autoref{sec:factorized_scheme} develops the diagrammatic proper-time prescription used for both scalar and fermion towers, with particular importance for the fermion tower where a direct heat-kernel treatment is less straightforward. \autoref{sec:result} presents the four-dimensional scalar and fermion results. \autoref{sec:discussion} summarizes the results and discusses their implications. The appendices collect the model definitions and compactification conventions, the cutoff-function analysis, the limitations of the heat-kernel treatment, the simplified string-tower extension, and technical details.

\section{Species scale}\label{sec:species}
In this section, we review the definitions of the species scale.
We first review the argument based on the perturbative computations~\cite{Dvali:2007wp,Castellano:2022bvr}. 
In this characterization, the species scale estimates the scale at which the perturbative description of gravity breaks down.
A parametric estimate is obtained by examining the one-loop correction to the graviton propagator in the presence of $N$ light particles.
These particles are referred to as species, and they are assumed to be weakly coupled only to gravity. This assumption is important when interpreting the gravitational loop correction as a universal species contribution, since additional interactions or operator mixing can obscure a simple running-coupling interpretation
\cite{Anber:2010uj,Anber:2011ut}.

By resumming the quantum correction corresponding to the vacuum-polarization correction to the gravitational coupling, we obtain the following expression for the inverse graviton propagator $G^{-1}(p^2)$ in four dimensions:
\begin{equation}
    \label{eq:grav_run}
    G^{-1}(p^2) \,\sim\, p^2 - \frac{N}{M_{\mathrm{pl},4}^{2}} \,p^4 \log\left(-\frac{p^2}{m^2}\right),
\end{equation}
where $M_{\mathrm{pl},4}$ is the four-dimensional Planck scale. The logarithmic running coupling is tied to the $1/\varepsilon$ divergence in dimensional regularization associated with curvature-squared terms~\cite{tHooft:1974toh}.
Comparing the loop and tree terms gives the parametric estimate
\begin{equation}
    \Lambda_{\mathrm{sp}} \sim \frac{M_{\mathrm{pl},4}}{\sqrt{N}},
\end{equation}
up to numerical factors and logarithmic corrections. This comparison serves as a consistency check within perturbation theory rather than a determination of the physical spectrum. Indeed, off-shell propagators are gauge dependent, and additional poles arising from a finite-order resummation of the self-energy need not be physical. A physical cutoff criterion requires further information, such as that supplied by scattering amplitudes~\cite{Caron-Huot:2024lbf}.

Recently, another definition of the species scale has been proposed~\cite{vandeHeisteeg:2022btw, vandeHeisteeg:2023ubh, vandeHeisteeg:2023dlw,Calderon-Infante:2025ldq}.
The second approach is to define the species scale as the suppression scale of higher-derivative corrections originating from quantum gravity. For example, in string theory, this corresponds to the $\alpha'$ expansion.  In this framework, the moduli-dependent species scale can be discussed. 
The higher-derivative corrections to the four-dimensional Einstein action coupled to a modulus $\phi$ are written as
\begin{equation}
    \label{eq:second_def}
    S = \frac{M_{\mathrm{pl},4}^2}{2} \int d^4x \sqrt{-g} \,\left( \mathcal{R} - (\nabla \phi)^2 + \sum_{n \geq 2} \frac{\mathcal{O}(\mathcal{R}^n)}{\Lambda_{\mathrm{sp}}^{2n-2}(\phi)} \right),
\end{equation}
where we assume the coefficient of each operator to be $\mathcal{O}(1)$. In a given EFT, the UV cutoff is typically identified with the mass scale of a new particle that the EFT does not describe. By integrating in such a new particle, one can extend the domain of validity of the EFT, thereby increasing the UV cutoff until additional new particles emerge. Although one might expect that iterating this procedure with an infinite tower of states would lead to an arbitrarily high UV cutoff, the iteration must terminate at a finite scale due to the intrinsic breakdown of the EFT description itself, as discussed in~\cite{vanBeest:2021lhn}.
The action Eq.~\eqref{eq:second_def} is the effective action of such a maximally extended EFT. The suppression scale $\Lambda_{\mathrm{sp}}(\phi)$ is identified as the scale where the EFT description breaks down.

\section{Scheme dependence and universality} \label{sec:universality}

We now analyze which parts of the tower-induced effective action are scheme dependent and which parts are stable under changes of the cutoff profile. This distinction is essential for relating the perturbative-breakdown and higher-derivative characterizations of the species scale, because the latter involves local EFT matching coefficients.

As we will see in the next section, the computation of the one-loop effective action from particles of mass $m$ involves the integration
\begin{align}
    \label{eq:qc_heat}
    I_{\text{1-loop}} &=\int_{\Lambda_{\mathrm{sp}}^{-2}}^{\infty} du \,u^{-1} e^{-um^2}
    =\int_{1}^{\infty} du \,u^{-1} e^{-um^2/\Lambda_{\mathrm{sp}}^2},
\end{align}
where we have introduced the species scale as the UV cutoff, and redefined the integration variable in the last equality. At this stage, $\Lambda_{\mathrm{sp}}$ is introduced only as a provisional UV cutoff for the loop integral; its identification with the species scale will be checked self-consistently from perturbative breakdown below (see Eq.~\eqref{eq:perturbative_check}). In the limit $(m/ \Lambda_{\mathrm{sp}}) \to 0$, we can expand this as
\begin{align}
    \label{eq:UV_expa}
    I_{\text{1-loop}} &= -\gamma + \log \frac{ \Lambda_{\mathrm{sp}}^2}{m^2} + \mathcal{O} \left(\frac{m^2}{ \Lambda_{\mathrm{sp}}^2}\right).
\end{align}
In quantum gravity, an infinite tower of states can appear~\cite{Ooguri:2006in}.
Let us consider the one-loop correction from integrating out an infinite tower of massive KK states, following~\cite{Cribiori:2023ffn, Aoufia:2024awo}.
The relevant integration is given by
\begin{equation}
    \mathcal{I}_{\text{1-loop}} = \sum_{n \geq 1} \int_{1}^{\infty} du \,u^{-1} e^{-u \cdot m_{\mathrm{KK}, n}^2/\Lambda_{\mathrm{sp}}^2}.
\end{equation}
Here $m_{\mathrm{KK}, n}^2=m_\mathrm{KK}^2 n^2$.
Due to the exponential suppression, the summation over the states is effectively truncated near $n=N \coloneqq \Lambda_{\mathrm{sp}}/m_{\mathrm{KK}}$.
Keeping only the logarithmic term in Eq.~\eqref{eq:UV_expa} and sharply truncating the tower gives, for large $N$,
\begin{align}
    \label{eq:qc_N}
    \sum_{n = 1}^{\lfloor N \rfloor} \log \frac{ \Lambda_{\mathrm{sp}}^2}{m_{\mathrm{KK},n}^2}
    = 2\int_{m_{\mathrm{KK}}}^{\Lambda_{\mathrm{sp}}} \frac{dq}{q}\,
    \left\lfloor\frac{q}{m_{\mathrm{KK}}}\right\rfloor = 2N - \log N + \mathcal{O}(1).
\end{align}
The floor function counts the KK levels lighter than the RG scale $q$. For a fixed curvature-squared operator in four dimensions, each logarithm carries the same single-mode coefficient, determined by its one-loop logarithmic UV divergence. Equation~\eqref{eq:qc_N} therefore has the usual interpretation as integrated one-loop RG evolution, with each mode's contribution to the beta function included above its mass threshold~\cite{Manohar:2018aog}. Once the massive modes have been integrated out, this sum contributes to a local threshold correction. It is distinct from the subsequent running of the low-energy EFT and from a nonanalytic momentum-space logarithm such as $\log(-p^2)$.

The leading term is proportional to the effective number of KK species $N$, and the logarithmic correction $\log N$ appears additively rather than multiplicatively. Because Eq.~\eqref{eq:qc_N} retains only the single-mode logarithms and sharply truncates the tower, its coefficient of $N$ is not the coefficient in the full proper-time calculation. Finite threshold terms and changes in the treatment of modes near the cutoff also contribute at order $N$. The full, untruncated calculation is presented in~\autoref{ssec:ndep}. The important point is that the computation so far assumes the specific regularization scheme.
Instead, we can also consider the generalized cutoff function:
\begin{equation}
    \label{eq:general_cutoff}
    \mathcal{I}_{\text{1-loop}}^{f_c} = \sum_{n\geq1} \int_{0}^{\infty} du \,u^{-1} e^{-u \cdot n^2/N^2(\phi_0)} \,f_c(u),
\end{equation}
where $f_c(u)$ is a cutoff function that regularizes the UV divergence, and satisfies
\begin{equation}
    \label{eq:bdy_cut_body}
    f_c(u) \to \left\{
    \begin{array}{ll}
    1 & \quad (u \to \infty), \\
    0 & \quad (u \to 0).
    \end{array}
    \right.
\end{equation}
Eq.~\eqref{eq:qc_heat} corresponds to the case $f_c(u) = H(u-1)$, where $H(x)$ is the Heaviside step function.
For the cutoff profiles satisfying the conditions in Appendix~\ref{app:uni_reg}, Eq.~\eqref{eq:general_cutoff} is computed as
\begin{align}
    \mathcal{I}_{\text{1-loop}}^{f_c}
    &=\frac{\sqrt{\pi}}{2} \widetilde{f}_c(-1/2) N(\phi_0) + \mathrm{Res}_{s=0} \widetilde{f}_c \cdot \log N(\phi_0) + \mathcal{O}(N^0(\phi_0)),
    \label{eq:one-loop_general_cut}
\end{align}
where $\widetilde{f}_c(s)$ is the Mellin transform of $f_c(u)$:
\begin{align}
    \widetilde{f}_c(s) := \int_{0}^{\infty} du \,u^{s-1} f_c(u),
\end{align}
and $\mathrm{Res}_{s=0} \widetilde{f}_c = -1$ due to the boundary condition Eq.~\eqref{eq:bdy_cut_body}. At the boundary of the moduli space, $N$ is large and the leading term is proportional to $N$. We see that the leading term is non-universal, as its numerical factor depends on the choice of regularization scheme. For instance, let us consider the cutoff function
\begin{equation}
    f_c(u) = 1-e^{-u}+\alpha u e^{-u^2},
\end{equation}
where $\alpha$ is a constant parameter. The Mellin transform is given by
\begin{align}
    \label{eq:cutoff_dependence}
    \widetilde{f}_c(-1/2) = 2\sqrt{\pi}+\frac{\alpha}{2}\Gamma(1/4).
\end{align}
This shows that the leading coefficient is sensitive to the choice of $\alpha$. The logarithmic coefficient, on the other hand, is unchanged within the cutoff class considered here. Its spin- and operator-dependent factor is inherited from the single-mode logarithmic UV divergence, in accord with the RG interpretation of Eq.~\eqref{eq:qc_N}.

Although the one-loop correction is scheme dependent, physical observables should not depend on the scheme (after performing field redefinitions properly), as it is related to the physical observables such as the scattering amplitudes.
This indicates that the scheme dependence of the one-loop correction should be compensated by the bare term.
For example, if we consider the $\mathcal{R}^2$ term, the structure of the bare term and the one-loop correction is given by
\begin{align}
    S_{\text{bare}} &= \int d^4 x \sqrt{-g} \,\mathcal{R}^2 \bigg(-\frac{\sqrt{\pi}}{2}\widetilde{f}_c(-1/2) +\text{(scheme-independent term)}\bigg) N(\phi_0) +\cdots ,\\
    S_{\text{1-loop}} &= \int d^4 x \sqrt{-g} \,\mathcal{R}^2 \,\mathcal{I}_{\text{1-loop}}^{f_c}. 
\end{align}
This provides a connection between the perturbative-breakdown and higher-derivative characterizations of the species scale. The former probes the loop correction to the graviton propagator, which is associated with curvature-squared terms, whereas the latter involves the corresponding local EFT coefficient. Since this coefficient includes both the loop contribution and the renormalized matching contribution, its numerical value is not fixed by the isolated loop calculation alone. Under the assumption that $\widetilde{f}_c(-1/2)$ is of order unity and that the renormalized matching contribution neither parametrically dominates nor cancels the tower term, the two characterizations are parametrically consistent.

The logarithmic coefficient is stable within the cutoff class considered here, while the coefficient of the leading local term depends on the cutoff profile. We therefore keep these two contributions separate in the explicit calculations for KK scalar and fermion towers that follow.

\section{One-loop effective action from a tower of KK scalars}
\label{sec:one_loop_scalar}

\subsection{Setup and heat kernel method}
We begin with the $d$-dimensional version of the minimal scalar tower model, setting the photon background to zero. We choose the mass spectrum and radion dependence of a real massless scalar on a circle $S^1$ of radius $R$. The Einstein-frame action is
\begin{align}
    &S = \frac{M_{\mathrm{pl}, d}^{d-2}}{2} \int d^d x \sqrt{-g} \Bigg( \mathcal{R}_d - ( \nabla \phi)^2 - \sum_{n \in \,\mathbb{Z}} \big( |\nabla \varphi_n|^2 + m_{\mathrm{KK}, n}^2 (\phi)  |\varphi_n|^2 \big) \Bigg),
\end{align}
where $\varphi_n$ denotes the KK scalar. The KK mass $m_{\mathrm{KK},n} (\phi)$ is given by
\begin{align}
    m_{\mathrm{KK},n} (\phi) = \frac{n}{R} \left( \frac{R_*}{R} \right)^{\frac{1}{d-2}}, \quad
    R = R_* \exp \left( - \sqrt{\frac{d-2}{d-1}} (\phi - \phi_*) \right),
\end{align} 
where $R_*$ represents the fixed vacuum expectation value (vev) of $R$. We follow the notation presented in Appendix~\ref{app:cc}. 

Let us now compute the one-loop contribution to the $d$-dimensional effective action obtained by integrating out the nonzero KK modes $\varphi_n \,(n \geq 1)$ using the heat kernel method~\cite{Gilkey:1995mj,Vassilevich:2003xt}. The corresponding one-loop effective action is given by
\footnote{We have used the relation $\varphi_{-n}^*=\varphi_n$.}%
\begin{align} 
    S_{\text{1-loop}} &= - \sum_{n \geq 1} \log \det (-\nabla^2 + m_{\mathrm{KK}, n}^2 (\phi)) = \int d^d x \sqrt{-g} \,\sum_{n \geq 1} \int_{0}^{\infty} \frac{du}{u} \,K(u, D_n).
    \label{eq:ol}
\end{align}
Here, $K(u, D_n)$ is a heat kernel defined as
\begin{equation}
    K(u, D_n) = \text{tr}_V e^{-u D_n}; \quad D_n \coloneqq - \nabla^2 + m_{\mathrm{KK}, n}^2 (\phi),
\end{equation}
and $V$ is a vector bundle on $d$-dimensional spacetime.
Assuming that the background fields vary relatively slowly, one can consider the adiabatic expansion of the heat kernel
\begin{equation} 
    \text{tr}_V e^{-u D_n} \simeq u^{-d/2} \sum_{k \geq 0} e_{2k}(x; D_n) u^k
\end{equation}
where the coefficient $e_{2k}(x; D_n)$ is a collection of the $2k$-dimensional operators~\cite{Vassilevich:2003xt,Avramidi:2000bm,Avramidi:1990je,Avramidi:1990ug}.

Although complete coefficients $e_{2k}(x; D_n)$ are explicitly available only to limited order, one can straightforwardly derive an expression for the one-loop action.  
By introducing the cutoff scale, we consider the following one-loop effective action:
\begin{equation} 
    \label{eq:ol_c}
    S_{\text{1-loop}} = \int d^d x \sqrt{-g} \,\sum_{n \geq 1} \int_{\Lambda_{\mathrm{sp}, 0}^{-2}}^{\infty} \frac{du}{u} \text{tr}_V e^{-u (- \nabla^2 + m_{\mathrm{KK}, n}^2 (\phi_0+\phi))},
\end{equation}
where $\phi$ represents a small fluctuation around the vev $\phi_0$.
Furthermore, we perform the derivative expansion of the effective action with respect to $|\nabla^2|/m_{\mathrm{KK}, n}^2 (\phi_0)$:
\begin{equation} \label{eq:ex_mass}
    \text{tr}_V e^{-u D_n} = u^{-d/2} e^{-u \cdot m_{\mathrm{KK}, n}^2 (\phi_0)} \sum_{k \geq 0} e_{2k}(x; \nabla, \widetilde{m}^2_{\mathrm{KK},n} (\phi)) u^k,
\end{equation}
where $\widetilde{m}^2_{\mathrm{KK},n} (\phi) \coloneqq m_{\mathrm{KK}, n}^2 (\phi_0+\phi) - m_{\mathrm{KK}, n}^2 (\phi_0)$, and $e_{2k}(x; \nabla, \widetilde{m}^2_{\mathrm{KK},n} (\phi))$ represents the part of $e_{2k}(x; D_n)$ that excludes terms containing only $\phi_0$. By substituting Eq.~\eqref{eq:ex_mass} into the one-loop action Eq.~\eqref{eq:ol_c}, one finds
\begin{align}
    S_{\text{1-loop}} &= \int d^d x \sqrt{-g} \,\sum_{n \geq 1, k \geq 0}
    e_{2k}(x; \nabla, \widetilde{m}^2_{\mathrm{KK},n} (\phi)) \int_{\Lambda_{\mathrm{sp},0}^{-2}}^{\infty} du \,u^{k-d/2-1}
    e^{-u \cdot m_{\mathrm{KK}, n}^2(\phi_0)} \\
    &=\int d^d x \sqrt{-g} \,\sum_{n \geq 1, k \geq 0}
    \,e_{2k}(x; \nabla, \widetilde{m}^2_{\mathrm{KK},n} (\phi)) \,\Lambda_{\mathrm{sp},0}^{d-2k} \int_{1}^{\infty} du \,u^{k-d/2-1} e^{-u \cdot n^2 / N^2(\phi_0)},
\end{align}
where we have introduced the dimensionless quantity $N(\phi_0) = \Lambda_{\mathrm{sp},0}/m_{\mathrm{KK}}(\phi_0)$, and changed the integration variable from $u$ to $\Lambda_{\mathrm{sp},0}^{-2}\,u$ in the second line.
To extract the coefficients at four-derivative order, we expand $e_{2k}(x; \nabla, \widetilde{m}^2_{\mathrm{KK},n} (\phi))$ in terms of the mass $m_{\mathrm{KK}, n}^{2j}(\phi_0)$:
\begin{equation}
    \label{eq:hc_split}
    e_{2k}(x; \nabla, \widetilde{m}^2_{\mathrm{KK},n} (\phi)) = \sum_{j \geq 0} e_{2k}^{\,(j)} (x; \nabla, \phi) m_{\mathrm{KK}, n}^{2j}(\phi_0).
\end{equation}
The heat-kernel coefficients ($k \leq 4$) relevant in this paper are summarized in \autoref{tab:heat-kernel-coeffs}. 

\begin{table*}[t]
\caption{Heat-kernel coefficients $e_{2k}(x;D)$ for $D = -\nabla^2 - E$, restricted to the terms relevant at four-derivative order~\cite{Vassilevich:2003xt}.}
\label{tab:heat-kernel-coeffs}
\centering
\renewcommand{\arraystretch}{1.4}
\begin{tabularx}{\textwidth}{@{}cX@{}}
\toprule
\textbf{Coefficient} & \textbf{Expression for $e_{2k}(x;D)$} \\
\midrule
$e_4^{(0)}$ &
$\displaystyle \frac{1}{(4\pi)^{d/2}} \left( 
\frac{1}{72} R^2 + 
\frac{1}{30} \Delta R - 
\frac{1}{180} R^{\mu \nu} R_{\mu \nu} + 
\frac{1}{180} R_{\mu \nu \rho \sigma} R^{\mu \nu \rho \sigma} \right)$ \\
\addlinespace
$e_6^{(1)}$ &
$\displaystyle \frac{1}{(4\pi)^{d/2}} \left(
\frac{1}{60} \Delta^2 E + 
\frac{1}{36} R \Delta E + 
\frac{1}{90} R^{\mu \nu} E_{; \mu \nu} + 
\frac{1}{30} R^{;\mu} E_{;\mu} + 
\frac{1}{30} E \Delta R \right)$ \\
\addlinespace
$e_8^{(2)}$ &
$\displaystyle \frac{1}{(4\pi)^{d/2}} \cdot \frac{1}{4!} \left(
\frac{2}{5} \Delta R E^2 + 
\frac{2}{3} R E \Delta E + 
\frac{1}{3} R (\nabla E)^2 + 
\frac{2}{3} R_{; \mu} E E^{; \mu} -
\frac{2}{15} R^{\mu \nu} E_{; \mu} E_{; \nu} \right.$ \\
&
$\displaystyle \quad
+ \frac{2}{15} E \Delta^2 E + 
\frac{8}{15} \Delta(E_{; \mu}) E^{; \mu} + 
\frac{4}{15} E (\Delta (E^{; \mu}))_{; \mu} + 
\frac{4}{15} (\Delta E)_{; \mu} E^{; \mu} + 
\frac{1}{3} (\Delta E)^2 $ \\
&
$\displaystyle \quad \left.
+ \frac{4}{15} E_{; \mu \nu} E^{; \mu \nu} \right)$ \\
\bottomrule
\end{tabularx}
\end{table*}


\subsection{One-loop effective action for a large number of species \texorpdfstring{$N$}{N}}
\label{ssec:ndep}

In this section, we examine the parametric behavior of the one-loop correction for $k=d/2$ (marginal), $k \leq d/2-1$ (relevant), and $k \geq d/2+1$ (irrelevant) operators in the decompactification limit. 
We show that the approximation performed in Eq.~\eqref{eq:qc_N} is consistent with a controlled large-\(N\) result. For simplicity, we first consider the $j=0$ part of the heat-kernel coefficients $e_{2k}^{(0)}(x; \nabla, \phi)$, followed by the result for the complete expression $e_{2k}(x; \nabla, \widetilde{m}^2_{\mathrm{KK},n} (\phi))$. We also restrict our attention to the even-dimensional case.

\vskip4pt

\paragraph{Correction to classically marginal operator \texorpdfstring{($k = d/2$)}{}}~\\[4pt]
We first consider the $k = d/2$ part of the one-loop correction. It reads
\begin{equation} \label{eq:ol_log}
    S_{\text{1-loop}}^{\,k= d/2} = \int d^d x \sqrt{-g} \,e_{d}^{(0)}(x; \nabla, \phi) \sum_{n \geq 1} \int_{1}^{\infty} du \,u^{-1} e^{-u \cdot n^2 / N^2(\phi_0)}.
\end{equation}
Taking the $N(\phi_0)$-derivative of the latter factor yields
\begin{equation}
    \label{eq:finite}
    \partial_{N(\phi_0)} \sum_{n \geq 1} \int_{1}^{\infty} du \,u^{-1} e^{-u \cdot n^2 / N^2(\phi_0)} = \frac{2}{N(\phi_0)} \sum_{n \geq 1} e^{-n^2 / N^2(\phi_0)}.
\end{equation}
By applying the Poisson resummation formula, we find
\begin{equation}
    S_{\text{1-loop}}^{\,k= d/2} = \int d^d x \sqrt{-g} \,e_{d}^{(0)}(x; \nabla, \phi) \left( \sqrt{\pi} N(\phi_0) - \log N(\phi_0) + \mathcal{O}(1,e^{-\pi^2 N^2(\phi_0)}) \right)
\end{equation}
in the decompactification limit $N(\phi_0) \to \infty$.

\vskip4pt

\paragraph{Corrections to classically relevant operator ($k \leq d/2-1$)}~\\[4pt]
The $k \leq d/2-1$ part of the one-loop correction is
\begin{equation} \label{eq:ol_low}
    S_{\text{1-loop}}^{\,k\leq d/2-1} = \int d^d x \sqrt{-g} \,e_{2k}^{(0)}(x; \nabla, \phi) \,\Lambda_{\mathrm{sp},0}^{d-2k} \int_{1}^{\infty} du \,u^{k-d/2-1} \sum_{n \geq 1} e^{-u \cdot n^2 / N^2(\phi_0)}.
\end{equation}
In the limit $N(\phi_0) \to \infty$, by applying the Poisson resummation formula directly to the summation over $n$, we find
\begin{equation}
    \begin{split}
    S_{\text{1-loop}}^{\,k\leq d/2-1} &\sim \int d^d x \sqrt{-g} \,e_{2k}^{(0)}(x; \nabla, \phi) \\
    &\qquad \times \left( \Lambda_{\mathrm{sp},0}^{d-2k} N(\phi_0) + \Lambda_{\mathrm{sp},0}^{d-2k} + \left( \frac{\Lambda_{\mathrm{sp},0}}{N(\phi_0)} \right)^{d-2k} + \mathcal{O}(e^{-\pi^2 N^2(\phi_0)}) \right).
    \end{split}
\end{equation}
These terms renormalize the action.
For example, the $k=0$ and $k=1$ terms in four dimensions renormalize the cosmological constant and Newton's constant, respectively. 

\vskip4pt

\paragraph{Corrections to classically irrelevant operator \texorpdfstring{($k \geq d/2+1$)}{}}~\\[4pt]
Finally, we consider the $k \geq d/2+1$ part of the one-loop correction:
\begin{equation}
    S_{\text{1-loop}}^{\,k \geq d/2+1} = \int d^d x \sqrt{-g} \,\sum_{n \geq 1, k \geq 0} e_{2k}^{(0)}(x; \nabla, \phi) \left( \frac{\Lambda_{\mathrm{sp},0}}{N(\phi_0)} \right)^{d-2k} n^{d-2k} \,\Gamma \left(k-\frac{d}{2}, \frac{n^2}{N^2(\phi_0)}\right),
\end{equation}
where $\Gamma(a,z)$ is the incomplete gamma function defined as $\Gamma(a,z) = \int_z^{\infty} dt \,t^{a-1} e^{-t}$.
By using the following representation for the incomplete gamma function
\begin{equation} \label{eq:rep_gam}
    \Gamma \left(k-\frac{d}{2}, z\right) = \Gamma\left(k-\frac{d}{2}\right) \,e^{-z} \sum_{l=0}^{\,k-d/2-1} \frac{z^l}{l!},
\end{equation}
the one-loop correction can be expressed as
\begin{align}
    \label{eq:ol_high}
    S_{\text{1-loop}}^{\,k \geq d/2+1} &= \int d^d x \sqrt{-g} \,e_{2k}^{(0)}(x; \nabla, \phi)
    \Gamma\left(k-\frac{d}{2}\right) \,\left( \frac{\Lambda_{\mathrm{sp},0}}{N(\phi_0)} \right)^{d-2k} \notag\\
    &\qquad \times \sum_{l=0}^{\,k-d/2-1} \frac{N^{-2l}(\phi_0)}{l!} \sum_{n \geq 1} n^{2(l-k+d/2)} e^{- n^2/N^2(\phi_0)}. 
\end{align}
Taking the $N^{-2}(\phi_0)$-derivative of the latter factor, one finds
\begin{align}
    &-\partial_{N^{-2}(\phi_0)}
    \sum_{l=0}^{\,k-d/2-1} \frac{N^{-2l}(\phi_0)}{l!} \sum_{n \geq 1} n^{2(l-k+d/2)}
    \,e^{- n^2 / N^2(\phi_0)} \notag\\
    &= \frac{N^{-2(k-d/2-1)}(\phi_0)}{(k-d/2-1)!}
    \sum_{n \geq 1} e^{-n^2 / N^2(\phi_0)} \sim N^{-2(k-d/2-3/2)}(\phi_0),
\end{align}
which gives a formula analogous to Eq.~\eqref{eq:finite}. Consequently, the one-loop correction Eq.~\eqref{eq:ol_high} behaves parametrically as
\begin{equation}
    S_{\text{1-loop}}^{\,k\geq d/2+1} \sim \int d^d x \sqrt{-g} \,e_{2k}^{(0)}(x; \nabla, \phi) \left( \left( \frac{N(\phi_0)}{\Lambda_{\mathrm{sp},0}} \right)^{2k-d} + \frac{N(\phi_0)}{\Lambda_{\mathrm{sp},0}^{2k-d}} \right).
\end{equation}
In this case, the first term is dominant, while the second term appears as the leading subdominant contribution. However, the first term merely represents the trivial breakdown of the EFT controlled by the KK mass gap, and can be ignored when considering the breakdown of the EFT description itself upon integrating in the KK states up to the species scale.

\vskip4pt

\paragraph{Species scale formula from one-loop correction}~\\[4pt]
Summarizing the results, the one-loop correction is expressed as
\begin{align}
    \label{eq:ol_num}
    S_{\text{1-loop}} &= S_{\text{1-loop}}^{\,k\leq d/2-1} + S_{\text{1-loop}}^{\,k= d/2} + S_{\text{1-loop}}^{\,k \geq d/2+1} \notag\\
    &\simeq \int d^d x \sqrt{-g} \,\sum_{k} e_{2k}^{(0)}(x; \nabla, \phi) \left( \left( \frac{N(\phi_0)}{\Lambda_{\mathrm{sp},0}} \right)^{2k-d} + \frac{N(\phi_0)}{\Lambda_{\mathrm{sp},0}^{2k-d}} \right).
\end{align}
The dimensionless quantity $N(\phi_0) \coloneqq \Lambda_{\mathrm{sp},0}/m_{\mathrm{KK}}(\phi_0)$ serves as a cutoff for the summation over KK modes $n$ as evident from the finiteness of the one-loop correction:
\begin{equation}
    \label{eq:tower_fin}
    \sum_{n \geq 1} e^{-n^2/N^2(\phi_0)}.
\end{equation}
Thus, this indeed shows that the KK modes below the species scale are being counted.

Furthermore, as explained before, if we take the energy scale of the theory to be the species scale and integrate in the KK states up to that scale, the first term in Eq.~\eqref{eq:ol_num} no longer controls the perturbative breakdown of the EFT. Therefore, the condition for the validity of the perturbative expansion is that the second term at $k=2$ be sufficiently small compared with the Einstein-Hilbert term; the higher-$k$ terms reproduce an equivalent condition. Treating this comparison as a parametric perturbativity estimate, under the matching assumptions stated in \autoref{sec:universality}, we obtain
\begin{equation}
    \label{eq:perturbative_check}
    \Lambda_{\mathrm{sp},0}^{2} \cdot \frac{1}{M_{\mathrm{pl}, d}^{d-2}} \cdot \frac{N(\phi_0)}{\Lambda_{\mathrm{sp},0}^{4-d}} \sim 1.
\end{equation}
This condition reproduces the standard species scaling
\begin{equation}
    \Lambda_{\mathrm{sp},0} \sim \frac{M_{\mathrm{pl}, d}}{N(\phi_0)^{\frac{1}{d-2}}}.
\end{equation}

Moreover, although the analysis relies on several assumptions, an analogous computation can also be carried out for a string tower, leading to an expression for the corresponding species scale. This is discussed in detail in Appendix~\ref{app:string_tower}.

\vskip4pt

\paragraph{General arguments}~\\[4pt]
Let us determine the dominant contribution for the full heat-kernel coefficients \\
$e_{2k}(x; \nabla, \widetilde{m}^2_{\mathrm{KK},n} (\phi))$ split as Eq.~\eqref{eq:hc_split}:
\begin{equation}
    e_{2k}(x; \nabla, \widetilde{m}^2_{\mathrm{KK},n} (\phi)) = \sum_{j \geq 0} e_{2k}^{\,(j)} (x; \nabla, \phi) m_{\mathrm{KK}, n}^{2j}(\phi_0).
\end{equation}
For $k \geq d/2+1$ contribution, the one-loop formula can be expressed as
\begin{align}
    S_{\text{1-loop}}^{\,k\geq d/2+1} &= \int d^d x \sqrt{-g} \sum_{j \geq 0}
    e_{2k}^{\,(j)} (x; \nabla, \phi) \Gamma\left(k-\frac{d}{2}\right) \left( \frac{N(\phi_0)}{\Lambda_{\mathrm{sp},0}} \right)^{2k-d-2j} \notag\\
    &\qquad \times \sum_{l=0}^{\,k-d/2-1} \frac{N(\phi_0)^{-2l}}{l!}
    \sum_{n \geq 1} n^{2(j+l-k+d/2)} e^{- n^2 / N^2(\phi_0)}.
\end{align}
Performing the same argument around Eq.~\eqref{eq:ol_high}, one obtains
\begin{equation}
    \label{eq:fol_num}
    S_{\text{1-loop}}^{\,k\geq d/2+1} \simeq \int d^d x \sqrt{-g} \sum_{j \geq 0} \bigg[ \left( \frac{N(\phi_0)}{\Lambda_{\mathrm{sp},0}} \right)^{2k-d-2j} +  \Lambda_{\mathrm{sp},0}^{2j} \,\frac{N(\phi_0)}{\Lambda_{\mathrm{sp},0}^{2k-d}} \bigg] \,e_{2k}^{\,(j)} (x; \nabla, \phi).
\end{equation}
Therefore, the $N$-dependence remains the same. One can show that the same is true for $k \leq d/2$. 


\section{Feynman diagram approach with proper-time regularization}\label{sec:factorized_scheme}

The preceding analysis has relied heavily on the heat kernel method. For KK fermions, however, a direct application of this method is not straightforward. In particular, a naive squaring of the Dirac operator can generate additional terms and, for a non-constant mass, may also lead to a multiplicative anomaly, as discussed in Appendix~\ref{app:heat_kernel}. By contrast, the Feynman-diagram approach remains directly applicable. We therefore use a diagrammatic prescription to define the local one-loop matching data. Regulator sensitivity in compactified and KK theories has been analyzed in~\cite{Branchina:2023rgi,Contino:2001gz}. These works motivate a careful treatment of tower sums, although they do not derive the specific factorized prescription adopted below.

The aim of this section is to demonstrate that, within the diagrammatic prescription adopted below, one-loop diagrams involving a KK tower can be organized into a factorized form consisting of a single-mode one-loop contribution and an effective KK-mode counting factor. This factorized form should be regarded as a calculational scheme for extracting local one-loop matching data. The scalar case provides a useful check, because the result can be compared with the heat-kernel computation in Sec.~\ref{sec:one_loop_scalar}. As discussed in Appendix~\ref{app:fact_ol}, the same type of expression can be derived for KK fermions. We note that, as mentioned in \autoref{sec:intro}, the leading local terms may depend on the scheme or prescription. The logarithmic coefficient, on the other hand, is the part that we treat as robust within the class of cutoff prescriptions considered here.

Motivated by the heat-kernel representation, we regulate the loop integrals using a proper-time cutoff. For a single propagator, this amounts to
\begin{align}
    \frac{1}{p^2+m_{\mathrm{KK}, n}^2 (\phi_0)}
    &= \int_{0}^{\infty} du \,e^{-u(p^2+m_{\mathrm{KK}, n}^2 (\phi_0))}
    \notag\\
    &\to \int_{\Lambda_{\mathrm{sp},0}^{-2}}^{\infty} du \,e^{-u(p^2+m_{\mathrm{KK}, n}^2 (\phi_0))}.
\end{align} 
More precisely, we first combine the propagators by introducing Feynman parameters, and then apply the proper-time regularization to the resulting denominator. In the following, we first discuss this method for the scalar case and subsequently extend the discussion to the fermion case.

\vskip4pt

\begin{figure}[htpb]
    \centering
    \includegraphics[width=0.3\linewidth]{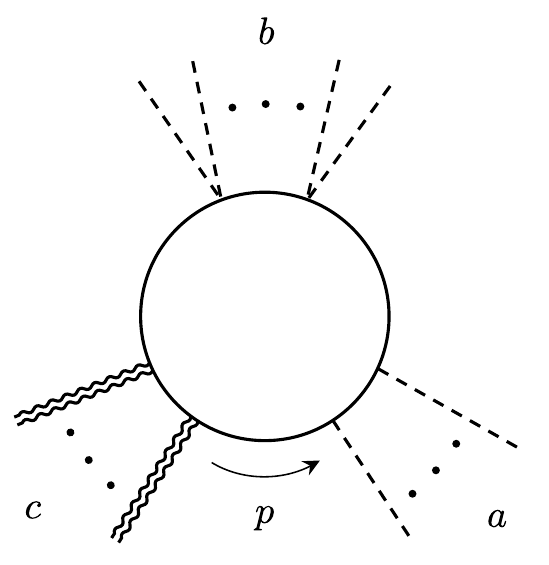}
    \caption{Generic one-loop diagram for KK scalar (solid line): This includes vertices with a single radion (dashed line), two radions, and a single graviton (double wavy line), with their respective numbers denoted by $a$, $b$, and $c$. The graviton-radion-(KK scalar)$^2$ vertex is not shown to avoid clutter; diagrams involving this vertex obey the same counting and are checked separately.}

    \par\smallskip
    \noindent\textbf{Alt text:}
    A schematic one-loop diagram with a closed solid line
representing a KK scalar. Three types of interaction
vertices are shown along the loop: a vertex with one
dashed radion line, a vertex with two dashed radion
lines, and a vertex with one double-wavy graviton
line. Their respective numbers are denoted by
$a$, $b$, and $c$.
    \label{fig:gene_ol}
\end{figure}

\paragraph{Demonstration of the proper-time regularization}~\\[4pt]
For simplicity, we restrict our analysis to the four-derivative corrections in the case of the KK scalar. The generic one-loop diagram is depicted in~\autoref{fig:gene_ol}.
According to the Feynman rules, the one-loop contribution is formally written as
\begin{equation}
    \label{eq:abc_ol}
    \mathcal{A}^{(a,b,c)}_n \sim m_{\mathrm{KK}, n}^{2(a+b)} (\phi_0)
    \int \frac{d^4p}{(2\pi)^4} \,\frac{1}{p^2+m_{\mathrm{KK}, n}^2 (\phi_0)} 
    \frac{\big( p^2 + p \cdot k + k^2 + m_{\mathrm{KK}, n}^2 (\phi_0) \big)^c}
    {\prod_{I=1}^{a+b+c-1} [(p+k_I)^2 + m_{\mathrm{KK}, n}^2 (\phi_0)]}.
\end{equation}

We define a hybrid procedure, combining Pauli--Villars and proper-time regularization, so as to reproduce the expression obtained via the heat kernel method in the scalar case; see Appendix~\ref{app:fact_ol} for details. First, to eliminate the superficial momentum divergences in the one-loop contribution Eq.~\eqref{eq:abc_ol}, we introduce an appropriate number of Pauli--Villars regulators with an auxiliary UV cutoff $\Lambda_{\mathrm{aux}}$. This cutoff is only an auxiliary regulator for the momentum integral and is not identified with the physical cutoff of the KK effective theory. As discussed in~\autoref{sec:universality}, the single-mode expression should be applied to the tower only for KK masses sufficiently below the species scale. We therefore do not take the limit $\Lambda_{\mathrm{aux}}\to\infty$ before introducing the species scale as a true physical cutoff. Instead, after combining the propagators by Feynman parameters, we apply the proper-time cutoff to the Pauli--Villars-regulated denominator. With this ordering, the auxiliary cutoff dependence drops out, and the result can be written as a finite single-mode contribution denoted by $\mathcal{A}_{\mathrm{single}}^{(a,b,c)}$ multiplied by the $n$-dependent factor. For contributions to marginal operators, however, $\mathcal{A}_{\mathrm{single}}^{(a,b,c)}$ is understood as the finite
coefficient obtained after subtracting the divergence originating from $\Lambda_{\mathrm{aux}}$, equivalently dropping the auxiliary pole factor $1/\varepsilon_{\mathrm{aux}}$ in the language of dimensional regularization:
\begin{equation}
    \label{eq:sfactor}
    \mathcal{A}^{(a,b,c)}_n = \mathcal{A}^{(a,b,c)}_{\mathrm{single}} \cdot \frac{1}{\wGamma(a+b)} \left( \frac{n}{N(\phi_0)} \right)^{2(a+b)} \int_{1}^{\infty} du \,u^{a+b-1} e^{-un^2/N^2(\phi_0)},
\end{equation}
where 
\begin{equation}
    \wGamma (a+b) = \left\{
    \begin{array}{ll}
    1 & \quad (a+b = 0)\\
    \Gamma(a+b) & \quad (a+b \geq 1).
    \end{array}
    \right.
\end{equation}
It is also important to note that $\mathcal{A}_{\mathrm{single}}^{(a,b,c)}$ is independent of the KK modes $n$. In the heat kernel method, the one-loop correction is graded by the mass dimension $2k$, which in this case equals $2(a+b)+4$.%
\footnote{Precisely, the parameter $k$ is given by $k=3a/2+2b+c$. In the case of four derivative corrections, however, the variables are subject to the constraint $4=a+2b+2c$, which leads to the simplified expression $k=a+b+2$.}
The same factorized tower form in Eq.~\eqref{eq:sfactor} also holds for KK fermions. In that case the Dirac trace, the spin connection, and the fermionic sign modify the
single-mode coefficient $\mathcal{A}_{\mathrm{single}}^{(a,b,c)}$, but they do not change the effective KK-mode counting factor that controls the dependence on the KK level $n$.

Using the method discussed in \autoref{sec:one_loop_scalar}, the sum over the KK modes appearing in Eq.~\eqref{eq:sfactor} is given by
\begin{align}
    \label{eq:four_del_num}
    &\sum_{n\geq1} \frac{1}{\wGamma(a+b)}
    \left( \frac{n}{N(\phi_0)} \right)^{2(a+b)} \int_{1}^{\infty} du \,u^{a+b-1} e^{-un^2/N^2(\phi_0)} \notag\\
    &\qquad = g_{a+b} \,N(\phi_0)
    - \delta_{a+b,0} \log N(\phi_0) + \mathcal{O}(N^0(\phi_0), e^{-\pi^2N^2(\phi_0)}),
\end{align}
where the first term is the leading contribution, while the second term is the subdominant contribution. 
The coefficients $g_{a+b}$ are given by
\begin{equation}
    g_{a+b} = \frac{\Gamma(a+b+1/2)}{\wGamma(a+b)} 
\end{equation}
The coefficients $g_{a+b}$ encode the moment-weighted counting of infinite KK levels induced by the proper-time cutoff. They therefore differ from the coefficient obtained by assigning unit weight to a finite number of particles truncated by hand. Finally, we obtain
\begin{align}
    \sum_{n\geq1} \mathcal{A}^{(a,b,c)}_n = \mathcal{A}^{(a,b,c)}_{\mathrm{single}} \cdot \Big( g_{a+b} \,N(\phi_0) - \delta_{a+b,0} \log N(\phi_0) \Big).
\end{align}


\section{Four-dimensional higher-derivative results} \label{sec:result}

We now collect the four-dimensional four-derivative terms obtained by integrating out the massive modes of the minimal scalar and Dirac towers. The expressions below should be read as the matter-loop part of the EFT matching data for the models defined in Appendix~\ref{app:cc}. In particular, the leading terms proportional to $\sqrt{\pi} N(\phi_0)$ and the
logarithmic terms have different status: as discussed in
\autoref{sec:universality}, the former depend on the choice of
regularization scheme and have to be combined with the corresponding bare higher-derivative terms, while the latter give the universal one-loop contribution. We keep the photon-dependent operators in the displayed formulas for completeness, but the field-basis discussion below focuses on
the radion and graviton sectors.

\vskip4pt

\paragraph{KK scalar}~\\[4pt]
The scalar-tower contribution at four-derivative order is
\begin{align}
    \label{eq:srp_4del}
    S_{\text{1-loop}}^{\,\partial^4} &= \frac{1}{(4 \pi)^2} \int d^4 x \sqrt{-g} \bigg\{\sqrt{\pi} N(\phi_0) \bigg[  \frac{1}{120} \mathcal{R}^2 + \frac{1}{60} \mathcal{R}^{\mu \nu} \mathcal{R}_{\mu \nu}
    - \frac{1}{12} \mathcal{R} (\nabla \phi)^2 \notag\\
    &\quad - \frac{1}{30} \mathcal{R}^{\mu \nu} \phi_{; \mu} \phi_{; \nu} + \frac{1}{20} (\nabla \phi)^4 - \frac{1}{180} \mathcal{R}_{\mu \nu \rho \sigma} F^{\mu \nu} F^{\rho \sigma} - \frac{1}{90} \mathcal{R}_{\mu \nu} F^{\mu \rho} F\indices{^{\nu}_{\rho}} \notag\\
    &\quad - \frac{1}{72} \mathcal{R} F_{\mu \nu} F^{\mu \nu} + \frac{1}{40} F_{\mu \nu} F^{\mu \nu} (\nabla \phi)^2 + \frac{1}{15} F_{\mu \rho} F\indices{_{\nu}^{\rho}} \phi^{;\mu} \phi^{;\nu} + \frac{1}{288} \left( F_{\mu \nu} F^{\mu \nu} \right)^2 \notag\\
    &\quad + \frac{1}{360} F_{\mu \rho} F\indices{_{\nu}^{\rho}} F^{\mu \sigma} F\indices{^{\nu}_{\sigma}} \bigg] - \log N(\phi_0) \left[ \frac{1}{120} \mathcal{R}^2 + \frac{1}{60} \mathcal{R}^{\mu \nu} \mathcal{R}_{\mu \nu} \right] \bigg\},
\end{align}
where we have ignored the terms with undifferentiated radion. This result
exhibits the general structure derived in the preceding sections. The first
bracket gives the leading local contribution from the tower in the chosen
regularization scheme, whereas the logarithmic curvature-squared terms are
the universal part of the scalar one-loop correction. Thus Eq.~\eqref{eq:srp_4del} is not by itself a physical Wilson coefficient; it is the
loop contribution that must be matched together with the bare local
higher-derivative terms of the UV theory
\cite{Manohar:2018aog,Burgess:2020tbq}.

To display the radion and graviton sectors in a convenient basis, we organize
the effective action into the perturbative expansion
\begin{equation}
    S = \sum_{k, l \geq 1} \lambda^{k-1} \zeta^{l-1} S_{k, l}, \quad
    \lambda \coloneqq \frac{1}{M_{\mathrm{pl}, 4}^2/N(\phi_0)}, \quad
    \zeta \coloneqq \frac{1}{M_{\mathrm{pl}, 4}^2/\log N(\phi_0)}.
\end{equation}
This $k$ is the same one used so far. We have ignored the $k=0$ term because it
does not require a field redefinition, and $S_{1,1}$ represents the original
action. The displayed operator basis is field-redefinition dependent
\cite{Criado:2018sdb}. We consider the following field redefinition,
\begin{equation} 
    \label{eq:frd}
    g_{\mu \nu} \longrightarrow g_{\mu \nu} - \frac{2\sqrt{\pi}}{(4 \pi)^2} \delta_{\lambda} g_{\mu \nu} + \frac{2}{(4 \pi)^2} \delta_{\zeta} g_{\mu \nu},
\end{equation}
where 
\begin{gather}
    \delta_{\lambda} g_{\mu \nu} = \lambda \left( c_1 \mathcal{R} \,g_{\mu \nu} + c_2 \mathcal{R}_{\mu \nu} + c_3 (\nabla \phi)^2 g_{\mu \nu} + c_4 \phi_{; \mu} \phi_{; \nu} \right); \\
    c_1 = \frac{1}{60}, \quad c_2 = - \frac{1}{60}, \quad c_3 = - \frac{1}{12}, \quad c_4 = \frac{1}{60},\\
    \delta_{\zeta} g_{\mu \nu} = \zeta \left( d_1 \mathcal{R} \,g_{\mu \nu} + d_2 \mathcal{R}_{\mu \nu} + d_3 (\nabla \phi)^2 g_{\mu \nu} + d_4 \phi_{; \mu} \phi_{; \nu} \right); \\
    d_1 = d_3 = -d_2 = -d_4 = \frac{1}{60},
\end{gather}
and the constants $c_i$ and $d_i$ are chosen to cancel all terms except
$(\nabla \phi)^4$ in this basis. The remaining scalar contribution is
\begin{align}
    &S_{\text{1-loop}}^{\,\partial^4} = \frac{1}{(4 \pi)^2} \int d^4 x \sqrt{-g} \bigg[ \sqrt{\pi} N(\phi_0) \bigg( - \frac{1}{24} (\nabla \phi)^4 \bigg) - \log N(\phi_0) \cdot \frac{1}{40} (\nabla \phi)^4 \bigg].
\end{align}
This last expression is a coefficient in the chosen field basis for the
one-loop effective action. Its leading $\sqrt{\pi} N(\phi_0)$ part is still
scheme dependent, and the displayed sign should not be interpreted as a
standalone positivity-bound statement in the standard flat-space sense
\cite{Adams:2006sv}.

\vskip4pt

\paragraph{KK fermion}~\\[4pt]
As discussed in Appendix~\ref{app:one-loop}, we use the integer KK spectrum corresponding to periodic boundary conditions. The diagrammatic method gives the four-derivative matching data for the minimally coupled Dirac tower, with photon interactions specified by Eq.~\eqref{eq:minimal_tower_action}. The fermion-tower contribution is
\begin{align}
    \label{eq:frp_4del}
    S_{\text{1-loop}}^{\,\partial^4} &= \frac{1}{(4 \pi)^2} \int d^4 x \sqrt{-g} \,\bigg\{ \sqrt{\pi} N(\phi_0) \bigg[ - \frac{1}{30} \mathcal{R}^2 + \frac{1}{10} \mathcal{R}^{\mu \nu} \mathcal{R}_{\mu \nu} + \frac{1}{12} \mathcal{R} (\nabla \phi)^2 \notag\\
    &\quad - \frac{11}{30} \mathcal{R}^{\mu \nu} \phi_{; \mu} \phi_{; \nu} + \frac{7}{40} (\nabla \phi)^4 + \frac{1}{45} \mathcal{R}_{\mu \nu \rho \sigma} F^{\mu \nu} F^{\rho \sigma} - \frac{13}{45} \mathcal{R}_{\mu \nu} F^{\mu \rho} F\indices{^{\nu}_{\rho}} \notag\\
    &\quad + \frac{1}{18} \mathcal{R} F_{\mu \nu} F^{\mu \nu} + \frac{13}{15} F_{\mu \nu} F^{\mu \nu} (\nabla \phi)^2 - \frac{7}{30} F_{\mu \rho} F\indices{_{\nu}^{\rho}} \phi^{;\mu} \phi^{;\nu} + \frac{1}{45} \left( F_{\mu \nu} F^{\mu \nu} \right)^2 \notag\\
    &\quad + \frac{7}{180} \left(F_{\mu \nu} \widetilde{F}^{\mu \nu} \right)^2 \bigg] - \log N(\phi_0) \left[ - \frac{1}{30} \mathcal{R}^2 + \frac{1}{10} \mathcal{R}^{\mu \nu} \mathcal{R}_{\mu \nu} \right] \bigg\}.
\end{align}
Again the first bracket is the scheme-dependent leading local contribution,
whereas the logarithmic term gives the universal one-loop part. Applying the
corresponding field-basis choice to the radion and graviton sector gives
\begin{align}
    S_{\text{1-loop}}^{\,\partial^4} = \frac{1}{(4 \pi)^2} \int d^4 x \sqrt{-g} \,\bigg[ \sqrt{\pi} N(\phi_0) \bigg( - \frac{1}{24} (\nabla \phi)^4 \bigg) - \log N(\phi_0) \cdot \frac{1}{15} (\nabla \phi)^4 \bigg].
\end{align}

We do not use the field-redefined coefficients above as a direct positivity-bound test, since such a test would require the corresponding renormalized amplitude data. Rather, the point of the explicit four-dimensional computation is that the same tower-summed pattern appears for both spin choices after the proper-time sum and the field-basis choice
are carried out. The calculation therefore supplies concrete
four-derivative EFT matching data: it identifies the scheme-dependent local tower contribution, isolates the universal logarithmic contribution, and shows how these two pieces are organized in the radion and graviton sectors
of the four-dimensional tower models.

\section{Summary and discussion}\label{sec:discussion}

We have computed the one-loop four-derivative effective action generated by the nonzero massive modes of minimally coupled scalar and Dirac towers with KK spectra in four dimensions. We used a hybrid diagrammatic proper-time scheme in which Pauli--Villars regularization defines each single-mode Feynman diagram, while a proper-time cutoff regulates the infinite KK tower sum.
This factorized framework treats scalar and fermion towers in parallel. The resulting four-dimensional matching data include operators involving the graviton, radion, and KK photon, with terms containing an undifferentiated radion omitted. Only the matter towers are integrated out. Massive gravitational KK modes are outside the chosen field content, and fermion Pauli couplings are set to zero in the defining action. The quantity $N(\phi_0)$ denotes the regulated number of levels of the matter tower at a constant radion background and does not include additional spin, polarization, charge, or internal degeneracy factors.

The main result is the separation of the leading and logarithmic contributions to the large-$N(\phi_0)$ tower threshold, as summarized in Eqs.~\eqref{eq:tower_cutoff_structure} and~\eqref{eq:one-loop_general_cut}. For the class of well-behaved cutoff functions discussed in Appendix~\ref{app:uni_reg}, the leading local term is generically proportional to $N(\phi_0)$. Its normalization depends on the cutoff profile through the Mellin transform $\widetilde{f}_c(-1/2)$, which is common to the leading terms in the factorized prescription. By contrast, the coefficient of the subleading logarithmic contribution $\log N(\phi_0)$ in the curvature-squared sector is fixed within the analyzed cutoff class by the residue entering Eq.~\eqref{eq:one-loop_general_cut}, which follows from the asymptotic conditions on the cutoff function.

Equation~\eqref{eq:qc_N} relates the logarithmic term to ordinary one-loop RG evolution across the KK thresholds. The explicit tower calculation retains the full mass dependence of the thresholds, since the single-mode logarithmic approximation is not valid for the entire infinite tower. It establishes that the inferred logarithmic coefficient survives this sum within the stated cutoff class, while determining the accompanying cutoff-dependent leading term and the conditions needed for the large-$N$ expansion.

The logarithmic threshold may also be useful in studying the moduli dependence of the species scale. From the viewpoint of the Distance Conjecture, the more meaningful quantum-gravity imprint might not be the coefficient-level prefactor of the species scale, but rather the exponential rate along moduli space inherited from the tower mass gap. The regulator-stable logarithmic threshold identified here is naturally sensitive to precisely this information. This suggests that the logarithmic part of the higher-derivative EFT data can be a possible probe of the moduli dependence of the species scale, in particular its exponential rate along moduli space, provided that the undetermined prefactor and ultraviolet matching do not alter the leading asymptotic dependence.

The same logarithmic term also gives a useful point of contact with amplitude-based cutoff criteria. The appropriate comparison is with the low-energy expansion of physical amplitudes, including the remaining matching and loop contributions. A full finite-momentum calculation across the KK thresholds could then test when loop effects become large and how this relates to amplitude-based cutoff criteria~\cite{Caron-Huot:2024lbf}. Establishing this relation explicitly, and examining the symmetry properties of the factorized regulator prescription at the amplitude level, are natural directions for future work.

The explicit scalar and fermion results are given in Eqs.~\eqref{eq:srp_4del} and~\eqref{eq:frp_4del}. Both contain cutoff-profile dependent local terms enhanced by $N(\phi_0)$, spin-dependent logarithmic curvature-squared terms, and KK-photon operators. This common structure reflects the factorized form of the calculation: the regulated KK tower factor is shared by scalar and fermion towers within the adopted prescription, while the corresponding single-mode matching coefficients are spin dependent. In the field basis described in \autoref{sec:result}, the radion and graviton sectors can be reduced to terms involving $(\nabla\phi)^4$. The resulting coefficients remain one-loop matching data in a chosen field basis and should not be interpreted as standalone observables or positivity statements.

These local loop contributions must be combined with counterterms and renormalized matching coefficients at a specified scale. Scheme-independent amplitudes or correlation functions of consistently matched operators are obtained when loop matrix elements, Wilson coefficients, and field redefinitions are treated consistently
\cite{Manohar:2018aog,Donoghue:2022eay,Criado:2018sdb}.
The separation established here makes the cutoff dependence of the tower contribution explicit before it is combined with the remaining matching data.

Within these matter-tower models, the present calculation shows the parametric compatibility of the perturbative-breakdown and higher-derivative characterizations provided that the cutoff-profile dependent factor multiplying the leading $N(\phi_0)$ term is of order one, the renormalized matching contribution is not parametrically larger than the tower term, and no tuned cancellation removes the generic enhancement. Under these assumptions, the perturbative-breakdown condition in Eq.~\eqref{eq:perturbative_check} reproduces the standard species scaling in $d$ dimensions
\cite{Dvali:2007wp,Castellano:2022bvr}.
Fixing the numerical normalization of the full higher-derivative coefficient requires ultraviolet matching data.

The simplified string-tower analysis in Appendix~\ref{app:string_tower} provides another natural extension. Under the assumptions that all excitations can be treated as scalars and that only the asymptotic degeneracy is retained, the scale in Eq.~\eqref{eq:species_scale_string} differs by a factor of $(\log N)^{1/2}$ from the scale inferred by naive cumulative state counting, while being parametrically consistent with the breakdown scale of the string tree-level expansion~\cite{Mende:1989wt,Basile:2023blg}.
This result is only indicative because a physical string spectrum also involves spin dependence, gauge constraints, ghosts, possible supersymmetric cancellations, winding sectors, and modular completion. Determining how these ingredients modify the weighted tower sum is a natural string-theoretic extension~\cite{Lee:2019wij,Antoniadis:1992sa,Green:1999pv}.

In summary, the KK calculation identifies a generic tower enhancement of local four-derivative terms and isolates spin-dependent logarithmic coefficients that are invariant within the stated proper-time cutoff class. Together, these results distinguish the normalization entering local species-scale estimates from the spin and KK mass-gap dependence retained by the logarithmic thresholds. The latter provide common data for comparing scalar and fermion towers across cutoff prescriptions and for studying their higher-derivative contributions as the KK mass gap varies over moduli space.


\section*{Acknowledgments}
I am deeply grateful to Yuta Hamada for extensive and valuable discussions, patient guidance, and careful reading of the manuscript. I thank Satoshi Iso for arranging a seminar where parts of this work were presented and for useful discussions. I am also grateful to Masamichi Miyaji for many helpful discussions at YITP. I would like to thank Shinji Mukohyama for continuous support and encouragement during the preparation and submission of this manuscript. I thank Kazumasa Okabayashi, Kei-ichiro Kubota, and Takafumi Kakehi for useful discussions. This work was supported by JST SPRING, Grant Number JPMJSP2110.



\appendix

\section{Minimal tower models and \texorpdfstring{$S^1$}{circle} compactification} \label{app:cc}

In this appendix, we specify the minimal tower models and introduce the circle-compactification conventions that motivate their spectra and radion dependence. We keep $d$ general and specialize to $d=4$ for the explicit results in the main text. We begin by specifying the sign conventions adopted in this paper. The Lorentzian metric is given a mostly positive signature, and the Euclidean metric is positive definite. The Riemann tensor is defined as $\mathcal{R}\indices{^{\mu}_{\nu \rho \sigma}} \coloneqq \Gamma\indices{^{\mu}_{\nu \sigma, \rho}} - \Gamma\indices{^{\mu}_{\nu \rho, \sigma}} + \cdots$, and the Ricci tensor is defined as $\mathcal{R}_{\mu \nu} \coloneqq \mathcal{R}\indices{^{\rho}_{\mu \rho \nu}}$.

To motivate the lower-dimensional models, consider an $S^1$ compactification of $D=d+1$-dimensional Einstein gravity coupled to a real massless scalar field $\Phi$ and a massless Dirac field $\Psi$. The higher-dimensional reference action is
\begin{equation}
    S = \frac{M_{\mathrm{pl}, D}^{D-2}}{2} \int d^D x \sqrt{-G} \,\left( \mathcal{R}_D - (\partial \Phi)^2 - \oPsi \slasholr{\nabla}  \Psi \right),
\end{equation}
where $M_{\mathrm{pl}, D}$ is a $D$-dimensional Planck scale. The antisymmetric derivative is defined by
\begin{gather}
    \oPsi \slasholr{\nabla} \Psi \coloneqq \oPsi \,\slashed{\nabla}\, \Psi - \oPsi \slashol{\nabla} \Psi = \oPsi \gamma^M \nabla_M \Psi - \oPsi \,\ol{\nabla}_M \gamma^M \Psi, \\
    \nabla_{M} \Psi \coloneqq
    \left(\partial_{M} + \frac{1}{4} \omega\indices{_{M}^{\mA \mB}} \gamma_{\mA \mB}\right) \Psi, \quad
    \oPsi \,\ol{\nabla}_{M} \coloneqq
    \oPsi \left(\ol{\partial}_{M} - \frac{1}{4} \omega\indices{_{M}^{\mA \mB}} \gamma_{\mA \mB}\right),
\end{gather}
where $M$ is a $D$-dimensional spacetime index, $\mA$ is a $D$-dimensional local Lorentz index, and $\omega\indices{_{M}^{ab}}$ is the spin connection, which is antisymmetric in local Lorentz indices. We use the second-order formalism with a torsion-free connection for the coupling of the fermion to gravity. Thus, we do not consider the four-fermion term $(\oPsi \gamma_{MNL} \Psi)(\oPsi \gamma^{MNL} \Psi).$ Furthermore, for simplicity, we only consider the case where the spacetime dimension $D$ is odd so that the dimension of the representation of the Clifford algebra is unchanged after the $S^1$ compactification. However, in the absence of fermions, the results are applicable to the case where $D$ is even.

The compactified direction $x^d$ is periodically identified as
\begin{equation}
    x^d \, \sim \, x^d + 2 \pi l 
\end{equation}
where $l$ is some length scale in the theory. For the compactification background, we retain only the metric zero modes:
\begin{equation}
    ds^2 = G_{MN} dx^{M} dx^{N} = e^{2 \alpha (\phi - \phi_0)} g_{\mu \nu} dx^{\mu} dx^{\nu}
    + e^{2 \beta \phi} (dx^d + A_{\mu} dx^{\mu})^2,
\end{equation}
where $g_{\mu \nu}$ is a $d$-dimensional metric, $A \coloneqq A_{\mu} (x^{\mu}) dx^{\mu}$ is a U(1) one-form, and $\phi = \phi (x^{\mu}) $ is a radion. We retain the names radion and KK photon for these fields in the lower-dimensional models. The massive KK modes of the higher-dimensional metric are not included in their field content. The exponents $\alpha, \beta$ are given by
\begin{equation}
    \alpha = \frac{1}{\sqrt{(d-2)(d-1)}}, \quad \beta = - (d-2) \alpha.
\end{equation}
We require that the metric approaches the KK vacuum 
\begin{equation}
    ds^2 = G_{MN} dx^{M} dx^{N} = g_{\mu \nu} dx^{\mu} dx^{\nu} + (R_0/l)^2 (dx^d)^2.
\end{equation}
as we take $r \to \infty$. In this limit, the radion $\phi$ asymptotes to the vev $\phi_0$. The circumference of $S^1$ is given by
\begin{equation} \label{eq:def_lr}
    2 \pi R(x) \coloneqq \int_{0}^{2 \pi l} dx^d \, e^{\beta \phi} = 2 \pi l e^{\beta \phi}.
\end{equation}
Again, it approaches the vev $R_0$ asymptotically. Since the $x^d$-direction is periodically identified, we can consider the Fourier expansion of the scalar $\Phi$  
\begin{equation}
    \Phi (x^{\mu}, x^d) = \sum_{n \in \mathbb{Z}} \varphi_n (x) e^{i n x^d / l}.
\end{equation}
In the case of the fermion $\Psi$, $S^1$ identification can be either periodic ($s=0$) or anti-periodic ($s=1$):
\begin{equation}
    \Psi (x^d) = (-1)^s \Psi (x^d + 2 \pi l).
\end{equation}
Thus, the Fourier expansion is given by
\begin{equation}
    \Psi (x^{\mu}, x^d) = \sum_{p \,\in\, \mathbb{Z}+s/2} \psi_p \,e^{ipx^d/l}.
\end{equation}

Motivated by these spectra, we define the lower-dimensional model with minimal scalar and fermion couplings by
\begin{align}
    S &= \frac{M_{\mathrm{pl}, d}^{d-2}}{2} \int d^d x \sqrt{-g} \,\bigg[ \mathcal{R}_{d} - (\nabla \phi)^2 + \frac{1}{4 g^2(\phi)} F^{\mu \nu} F_{\mu \nu} \notag\\
    &\quad - \sum_{n \in \,\mathbb{Z}} \left( | D \varphi_n |^2 + m_{\mathrm{KK}, n}^2 (\phi) |\varphi_n|^2 \right) - \sum_{p \,\in\, \mathbb{Z}+s/2} \left( \opsi_p \slasholr{\mathcal{D}} \psi_p - 2 m_{\mathrm{KK}, p} (\phi) \,\opsi_p \psi_p \right) \bigg]. \label{eq:minimal_tower_action}
\end{align}
The nonminimal scalar curvature coupling and fermion Pauli couplings are set to zero in this defining matter action. The latter choice distinguishes the model from the full circle reduction of the higher-dimensional Dirac action: the torsion-free spin connection generates a Pauli interaction with the KK photon even when the higher-dimensional fermion is massless. We omit this interaction by definition, without assuming that its contribution to photon-dependent four-derivative terms is parametrically suppressed. At $A_\mu=0$, the Pauli interaction vanishes, so this choice does not affect the matter-loop graviton--radion sector. The lower-dimensional spin connection is retained in the covariant derivative below.

The masses $m_{KK,n} (\phi)$ and $m_{KK,p} (\phi)$ are defined by
\begin{equation}
    \begin{gathered}
    m_{KK,n} (\phi) \coloneqq n \cdot m_{KK} (\phi), \quad m_{KK,p} (\phi) \coloneqq p \cdot m_{KK} (\phi), \\
    m_{KK} (\phi) = \frac{1}{R} \left( \frac{R_0}{R} \right)^{\frac{1}{d-2}}.
    \end{gathered}
\end{equation}
The gauge coupling $g (\phi)$ is defined by
\begin{equation}
    g (\phi) \coloneqq \frac{1}{R_0} e^{ \sqrt{\frac{d-1}{d-2}} (\phi - \phi_0)} = m_{KK} (\phi).
\end{equation}
Finally, the covariant derivatives $D$ and $\mathcal{D}$ are given by
\begin{gather}
    D_{\mu} \varphi_n = ( \partial_{\mu} - i n A_{\mu} ) \varphi_n, \\
    \mathcal{D}_{\mu} \psi_p = \left( \nabla_{\mu} - i p  A_{\mu} \right) \psi_p; \quad \nabla_{\mu} \psi_p = ( \partial_{\mu} + \frac{1}{4} \omega\indices{_{\mu}^{ab}} \gamma_{ab} ) \psi_p,
\end{gather}
where $\omega\indices{_{\mu}^{ab}}$ is the $d$-dimensional torsion-free spin connection. The one-loop calculations integrate out the nonzero massive modes of the scalar or Dirac tower separately, keeping $g_{\mu\nu}$, $\phi$, and $A_\mu$ as background fields.


\section{Universality and non-universality of regularization schemes} \label{app:uni_reg}

In this appendix, we examine how the choice of regularization scheme influences the UV divergence and the truncation of the infinite tower of states. We have already discussed one particular choice, which gives Eq.~\eqref{eq:four_del_num}:
\begin{align}
    \label{eqn:sharp_cut}
    \mathcal{N}_{a+b} &\coloneqq \sum_{n\geq1} \frac{1}{\wGamma(a+b)}
    \left( \frac{n}{N(\phi_0)} \right)^{2(a+b)} \int_{1}^{\infty} du \,u^{a+b-1} e^{-un^2/N^2(\phi_0)} \notag\\
    &= g_{a+b} \,N(\phi_0)
    - \delta_{a+b,0} \log N(\phi_0) + \mathcal{O}(N^0(\phi_0), e^{-\pi^2 N^2(\phi_0)}).
\end{align}
We will show that the factor $g_{a+b}$ and $\delta_{a+b,0}$ themselves are independent of the choice of regularization scheme. Before giving the general argument, however, we give another derivation, illustrating the underlying structure for any cutoff profile. To do so, we first rewrite $\mathcal{N}_{a+b}$ as 
\begin{equation}
    \label{eqn:step_ol}
    \mathcal{N}_{a+b} = \sum_{n\geq1} \frac{1}{\wGamma(a+b)}
    \left( \frac{n}{N(\phi_0)} \right)^{2(a+b)} \int_{0}^{\infty} du \,u^{a+b-1}
    e^{-u \cdot n^2/N^2(\phi_0)} \,H(u-1),
\end{equation}
where $H(x)$ is the step function, and view the $u$-integral as the Mellin transform. The Mellin transform of the exponential term is given by
\begin{equation}
    \mathcal{M} \left[ \sum_{n\geq1} n^{2(a+b)}
    e^{-u \cdot n^2/N^2(\phi_0)} \right] (s) = N^{2s} \zeta(2(s-(a+b))) \Gamma(s),
\end{equation}
with the fundamental strip $\mathrm{Re}(s)>a+b+1/2$. While the Mellin transform of the step function is given by
\begin{equation}
    \label{eq:step_mel}
    \mathcal{M} \left[ H(u-1) \right] (s) = - \frac{1}{s},
\end{equation}
where the fundamental strip is $\mathrm{Re}(s)<0$. Therefore, for some constant $c > a+b+1/2$, Eq.~\eqref{eqn:step_ol} can be written as
\begin{equation}
    \label{eq:mel_convol}
    \frac{1}{\wGamma(a+b)} \int_{c-i \infty}^{c+i\infty} \frac{ds}{2 \pi i}
    N^{2(s-(a+b))} \zeta(2(s-(a+b)))
    \Gamma(s) \,\frac{1}{s-(a+b)}.
\end{equation}
Thus, the leading term can be obtained from the residue at $s=a+b+1/2$, while the subleading term appears at $s=a+b=0$. 

To clarify what really happens, we consider another cutoff function deviating from the step function:
\begin{equation}
    \frac{1}{2} \left( 1 + \tanh{\left(\frac{l}{2} \log u \right)} \right) = \frac{u^l}{1+u^l},
\end{equation}
with $l>0$. The Mellin transform of this function is given by
\begin{align}
    \mathcal{M} \left[ \frac{u^l}{1+u^l} \right] (s)
    &= - \frac{1}{l} B \left( \frac{s}{l}, 1- \frac{s}{l} \right) = \frac{-\pi/l}{\sin (\pi s/l)} \\
    &= - \frac{1}{s} - \frac{\pi^2}{6l^2}s + \mathcal{O}(s^2),
    \label{eqn:logi_mel}
\end{align}
with the fundamental strip $-l < \mathrm{Re}(s) <0$. Thus, we find that the pole structure at $s=0$ is identical to that of the step function Eq.~\eqref{eq:step_mel}. However, care must be taken regarding the conditions under which the constant $c$ exists. It reads $a+b+1/2<c<a+b+l$, implying that $l>1/2$ is the condition for the cutoff function to be well-behaved. Since the Mellin transform Eq.~\eqref{eqn:logi_mel} at $s=-1/2$ depends on the free parameter $l$, the coefficient of the leading term is non-universal. However, we still have $g_{a+b}$ as is evident from $\Gamma(s)$ in Eq.~\eqref{eq:mel_convol}, and, thus, this might be universal.

\vskip4pt

\paragraph{General arguments}~\\[4pt]
Let us show that the above guess is indeed valid in the factorized scheme for any well-behaved cutoff profile, with the precise meaning of "well-behaved" to be discussed below. The original improper integral is 
\begin{equation}
    \label{eq:improper}
    \sum_{n\geq1} \frac{1}{\wGamma(a+b)}
    \left( \frac{n}{N(\phi_0)} \right)^{2(a+b)} \int_{0}^{\infty} du \,u^{a+b-1} e^{-un^2/N^2(\phi_0)},
\end{equation}
whose integrand for each fixed KK mode has a logarithmic singularity at $u=0$ when $a+b=0$. By the structure theorem for point-supported distributions, extensions preserving the scaling degree have a one-dimensional local renormalization ambiguity~\cite{Ooguri:1984divergent, Hormander2003,Kanwal1998,GelfandShilov:1964}\footnote{The author gratefully acknowledges K. Okabayashi for valuable discussions concerning the paper~\cite{Ooguri:1984divergent}.}
. Regularization prescriptions that yield extensions preserving the scaling degree therefore give equivalent renormalized single-mode integrals up to a finite local subtraction. We may thus use proper-time cutoff regularization as a representative prescription and examine its profile dependence in the regulated infinite tower below.

Let $f_c(u)$ be a function which regularizes the improper integral Eq.~\eqref{eq:improper} as
\begin{equation}
    \mathcal{N}_{a+b} [f_c] \coloneqq \sum_{n\geq1} \frac{1}{\wGamma(a+b)}
    \left( \frac{n}{N(\phi_0)} \right)^{2(a+b)} \int_{0}^{\infty} du \,u^{a+b-1}
    e^{-u \cdot n^2/N^2(\phi_0)} \,f_c(u),
\end{equation}
with the following boundary conditions:
\begin{equation}
    \label{eq:bdy_cut}
    f_c(u) \to \left\{
    \begin{array}{ll}
    1 & \quad (u \to \infty), \\
    0 & \quad (u \to 0).
    \end{array}
    \right.
\end{equation}
We denote the Mellin transform of $f_c(u)$ as $\widetilde{f}_c(s) \coloneqq \mathcal{M}[f_c](s)$. Its pole structure is completely determined by the asymptotic behavior of the original function $f_c(u)$ at $u=0$ and $u=\infty$. In fact, in general, if $f(x)$ admits a power-logarithmic asymptotic expansion of the form
\begin{equation}
    f(x) \sim c_{\xi, \lambda}^{(0, \infty)} x^{\xi} (\log x)^{\lambda}
\end{equation}
as $x \to 0$ or $x \to \infty$, then its Mellin transform admits a corresponding singular expansion 
\begin{equation}
    \widetilde{f}(s) \sim \pm \,c_{\xi, \lambda}^{(0, \infty)} \frac{(-1)^{\lambda} \lambda!}{(s+\xi)^{\lambda+1}},
\end{equation}
for an appropriate fundamental strip~\cite{FLAJOLET19953}. The sign $\pm$ is positive for $x \to 0$ and negative for $x \to \infty$.

To control the contribution from the large-$u$ endpoint near $s=0$, we impose $f_c(u)-1=\mathcal{O}(u^{-\rho})$ as $u\to\infty$ for some $\rho>0$. In the power-logarithmic expansion above, a first correction $u^{-\alpha}(\log u)^\lambda$ with $\alpha>0$ and a remainder of smaller order satisfies this condition for any $0<\rho<\alpha$. Subtracting the constant asymptotic term gives
\begin{equation}
    \widetilde{f}_c(s) = -\frac{1}{s}
    + \int_0^1 du\,u^{s-1}f_c(u)
    + \int_1^\infty du\,u^{s-1}\bigl(f_c(u)-1\bigr).
\end{equation}
The last integral is holomorphic for $\mathrm{Re}(s)<\rho$, while the small-$u$ condition discussed below makes the first integral holomorphic near $s=0$. Thus, the condition at infinity provides a sufficient criterion for meromorphic continuation across $s=0$, where the normalization in Eq.~\eqref{eq:bdy_cut} produces an isolated simple pole with residue
\begin{equation}
    \textrm{Res}_{s=0} \,\widetilde{f}_c(s) = c_{0,0}^{(0)} - c_{0,0}^{(\infty)} = -1.
\end{equation}

We next consider the existence of the common strip with the zeta function. We restrict attention to cutoff profiles for which there exists a number $l_0 \in \mathbb{R}^+ \cup \{ \infty \}$ such that $u^{-l_0 + \varepsilon} f_c(u) \to 0$ for any $\varepsilon>0$ as $u \to 0$. Here, $l_0=\infty$ means that $u^{-l_0 + \varepsilon} f_c(u) \to 0$ is satisfied for any $l_0 \in \mathbb{R}^+$. Taking the supremum over allowed $l_0(\eqqcolon l_0^*)$, then the condition for the existence of the common strip is given by $l_0^*>1/2$, so that a contour with $a+b+1/2<c<a+b+l_0^*$ can be chosen. This also ensures that $c_{1/2,\lambda}^{(0)}=0$ for $f_c(u)$, and, thus, $|\widetilde{f}_c(s=-1/2)|<\infty$. Together, these two endpoint conditions allow the Mellin contour to be shifted across the poles that generate the leading and logarithmic terms. Therefore, for cutoff profiles satisfying these conditions, the improper integral Eq.~\eqref{eq:improper} is regularized as 
\begin{equation}
    \mathcal{N}_{a+b} [f_c] = \frac{1}{2} g_{a+b} \,\widetilde{f}_c(-1/2) N(\phi_0) + \delta_{a+b,0} \,\mathrm{Res}_{s=0} \widetilde{f}_c \cdot \log N(\phi_0) + \mathcal{O}(N^0(\phi_0)),
\end{equation}
where $|\widetilde{f}_c(-1/2)|<\infty$ and $\mathrm{Res}_{s=0} \widetilde{f}_c=-1$.

\vskip4pt

\paragraph{Remarks on the cutoff function}~\\[4pt]
In the preceding analysis, we employed a single, universal cutoff for all KK modes. More generally, one may introduce distinct cutoffs for each mode individually. The leading and logarithmic terms remain unchanged under the summability condition stated below.

Let $f_c^{(n)}(u)$ denote the cutoff function associated with the KK mode $n$. When the sum and the Mellin integral can be interchanged on a common contour, the result can be written as
\begin{equation}
    \mathcal{N}_{a+b} [(f_c^{(n)})_{n \geq 1}] = \frac{1}{\wGamma(a+b)} \int_{c - i \infty}^{c + i \infty} \frac{ds}{2 \pi i} \,N^{2(s-(a+b))}(\phi_0) \,\Gamma(s) \sum_{n \geq 1}
    \frac{\widetilde{f}_c^{\,(n)}(a+b-s)}{n^{2(s-(a+b))}}.
\end{equation}
We restrict this extension to fixed cutoff profiles satisfying
\begin{equation}
    \sum_{n\geq1} \int_0^\infty \frac{du}{u}\,
    \left|f_c^{(n)}(u)-f_c(u)\right| < \infty,
\end{equation}
where $f_c(u)$ is a common cutoff profile for which the preceding large-$N$ expansion holds. This condition prevents the differences between the mode-dependent and common cutoffs from accumulating into an additional leading or logarithmic term. Indeed, with the measure written as $du/u$, the cutoff difference in each mode is multiplied by $(un^2/N^2(\phi_0))^{a+b}e^{-un^2/N^2(\phi_0)}/\wGamma(a+b)$. For fixed $a+b\geq0$, this factor is bounded independently of $u$, $n$, and $N(\phi_0)$, so the condition above limits the change in $\mathcal{N}_{a+b}$ to $\mathcal{O}(N^0(\phi_0))$. The coefficients of both $N(\phi_0)$ and $\log N(\phi_0)$ are therefore those of the common profile $f_c$.

\subsection{The exponential regularization (\texorpdfstring{$l_0=\infty$}{exponential decay at the origin})}

In this section, we examine in detail the case of exponential regularization $f_c^{\,\mathrm{exp}}(u) = \exp (-1/u)$ as an illustrative example with \(l_0 = \infty\). This analysis elucidates the essential aspects of the computation performed when determining the $N$-dependence of the one-loop correction. To begin, the Mellin transform of $f_c^{\,\mathrm{exp}}(u)$ can be expressed as
\begin{equation}
    \widetilde{f}_c^{\,\mathrm{exp}} (s) = \Gamma(-s) = - \frac{1}{s} \times \Gamma(1-s).
\end{equation}
As is evident from Eqs.~\eqref{eq:step_mel} and~\eqref{eqn:logi_mel}, the choice of regularization scheme corresponds precisely to how the factor $-1/s$ is dressed.  In particular, employing the Gamma function for this dressing is especially advantageous from a number-theoretic perspective. Indeed, the zeta function has already been completed in Eq.~\eqref{eq:mel_convol} when $a+b=0$. In fact, by defining $\Lambda(s) \coloneqq \pi^{-s/2} \Gamma(s/2) \zeta(s)$, called the completed Riemann zeta function, one directly obtains
\begin{equation}
    \mathcal{N}_{a+b=0} [H(u-1)] = \int_{c-i \infty}^{c+i\infty} \frac{ds}{2 \pi i} \pi^s N^{2s} s^{-1} \Lambda(2s).
\end{equation}
In this language, the automorphy of the theta function utilized in~\autoref{ssec:ndep} is precisely equivalent to the Hecke-type functional equation $\Lambda(2s)=\Lambda(1-2s)$~\cite{Bochner:1951}. This relation is nothing but the well-known correspondence between automorphic forms and their $L$-functions. In what follows, we will briefly see how other equivalent formulas~\cite{Kanemitsu:2002} appear in the step-function regularization and in the exponential regularization. First, there exists a representation in terms of the incomplete gamma function~\cite{Lavrik_1991, Kanemitsu_2002}
\begin{align}
    \sum_{n=1}^{\infty} n^{-2s} \Gamma (s, \pi n^2 w) = \Gamma(s) \zeta(2s) - \pi^{2s-\frac{1}{2}} \sum_{n=1}^{\infty} n^{2s-1} \Gamma\left(\frac{1}{2}-s, \frac{\pi n^2}{w}\right) - \frac{\pi^s w^{s- \frac{1}{2}}}{2s-1} + \frac{\pi^s w^s}{2s},
\end{align}
with $\mathrm{Re} \,w > 0$. Setting $s=k-d/2$ and $w=1/(\pi N^2(\phi_0))$, one recovers precisely the expression that appeared in the calculation of the $N$-dependence under the step-function regularization, providing its exact closed form. Next, one may also employ an alternative representation involving the modified $K$-Bessel function%
\footnote{As an aside, this expression characterizes the behavior of the Ramanujan–Guinand formula, which is known to be equivalent to the weak-coupling expansion of the non-holomorphic Eisenstein series~\cite{Berndt2008}.}%
\begin{equation}
    \label{eq:watson}
    \sum_{n=1}^{\infty} (\pi n z)^{\nu} K_{{\nu}} (2 \pi n z) = \frac{\Gamma(\nu + \frac{1}{2})}{4 \sqrt{\pi} z} - \frac{\Gamma(\nu)}{4} + \frac{z^{2\nu} \Gamma(\nu + \frac{1}{2})}{2 \sqrt{\pi}} \sum_{n=1}^{\infty} \frac{1}{(n^2 + z^2)^{\nu + 1/2}}.
\end{equation}
This formula is known as the Watson's formula~\cite{WATSON_1931, Elizalde2012ten} and corresponds directly to the expression that appears in the exponential regularization. In fact, it reads
\begin{align}
    \label{eq:exp_reg_K}
    \mathcal{N}_{a+b} [f_c^{\,\mathrm{exp}}]
    &= \sum_{n\geq1} \frac{1}{\wGamma(a+b)}
    \left( \frac{n}{N(\phi_0)} \right)^{a+b} \int_{0}^{\infty} du \,u^{a+b-1}
    \exp \left[ - \frac{n}{N(\phi_0)} \left( u + \frac{1}{u} \right) \right] \\
    &= \sum_{n\geq1} \frac{2}{\wGamma(a+b)}
    \left( \frac{n}{N(\phi_0)} \right)^{a+b} K_{a+b} \left( \frac{2n}{N(\phi_0)} \right).
\end{align}
This coincides with the Watson's formula when $\nu = a+b$ and $z = 1/(\pi N(\phi_0))$. In particular, considering the case $a + b = 0$, one obtains
\begin{align}
    \mathcal{N}_0 [f_c^{\,\mathrm{exp}}] &= 2 \sum_{n=1}^{\infty} K_0 \left( \frac{2n}{N(\phi_0)} \right) \\
    &= \frac{\pi}{2} N(\phi_0) - \log (2 \pi N(\phi_0)) + \gamma + \sum_{n=1}^{\infty}
    \left( \frac{1}{\sqrt{n^2 + (\pi N(\phi_0))^{-2}}} - \frac{1}{n} \right) \\
    &= \frac{\pi}{2} N(\phi_0) - \log (2 \pi N(\phi_0)) + \gamma + \sum_{m=1}^{\infty} \dbinom{-1/2}{m}
    \zeta (2m+1) (\pi N(\phi_0))^{-2m}.
\end{align}

\vskip4pt

\paragraph{Remarks on the modular-invariant completion}~\\[4pt]
Thus far, we have encountered three types of regularization: dimensional (or Pauli–Villars) regularization for Feynman diagrams, the proper-time method for handling an infinite tower of states, and exponential-type cutoffs for regulating UV divergences. Each of these admits a modular-invariant completion~\cite{Abel:2021tyt,Abel:2023hkk,Abel:2024twz}. 

The first corresponds to the insertion of a non-holomorphic Eisenstein series. The second bears resemblance to the modular-invariant partition function of string theory compactified on a circle, with the winding modes completely decoupled. The third belongs to the class of modular-invariant regulators, making the original integral invariant under the $S$-transformation $u \to 1/u$ for $a+b=0$, as evident from Eq.~\eqref{eq:exp_reg_K}.


\section{Limitations of the heat kernel method} \label{app:heat_kernel}

While this approach is powerful, its reliance on a second-order differential operator imposes certain limitations when extending the analysis to the case of the KK fermion. We now clarify how this issue arises in this context.

Consider the minimally coupled Dirac tower defined in Eq.~\eqref{eq:minimal_tower_action}, with the photon background set to zero. For simplicity in the calculations, we work on Euclidean spacetime~\cite{vanNieuwenhuizen:1996tv}. The four-dimensional action reads
\begin{align}
    \label{eq:ferm_ac}
    S = \frac{M_{\mathrm{pl}, 4}^2}{2} \int d^4 x \sqrt{g} \,\bigg[ \mathcal{R} + (\nabla \phi)^2 + \sum_{p \,\in\, \mathbb{Z}+s/2} \left( \opsi_p \slasholr{\mathcal{D}} \psi_p - 2 m_{\mathrm{KK}, p} (\phi) \,\opsi_p \psi_p \right) \bigg],
\end{align}
where $\opsi \coloneqq \psi^{\dagger} \gamma^{\ov{5}}$. Here, the barred index indicates a local `Euclidean' index. The covariant derivative is defined as
\begin{equation}
    \mathcal{D}_{\mu} \psi_p = (\partial_{\mu} + \frac{1}{4} \omega\indices{_{\mu}^{ab}} \gamma_{ab}) \psi_p.
\end{equation}
The parameter $s$ in the summation index indicates whether the fermion is periodic ($s = 0$) or anti-periodic ($s = 1$) on $S^1$. Considering the quantum fluctuation of the KK fermions $\psi_p$, where $p \in \mathbb{Z}^* \coloneqq (\mathbb{Z}+s/2) \setminus \{ 0 \}$, the one-loop effective action is given by
\begin{equation}
    S_{\text{1-loop}} = - \sum_{p \,\in \,\mathbb{Z}^*} \log \det \left( \slashed{\mathcal{D}} - m_{KK,p} (\phi_0+\phi) \right).
\end{equation}
This one-loop action can be computed using either the heat kernel method or the Feynman diagram approach. However, both methods exhibit certain limitations. 
We begin by discussing the heat kernel method. Since this method is not applicable to first-order operators, we typically employ the doubling trick~\cite{DeBerredo-Peixoto:2001wkv, Goncalves:2009sk, Karan:2017txu}:
\begin{align}
    S_{\text{1-loop}} &= - \sum_{p \,\in \,\mathbb{Z}^*} \log \det \left( \slashed{\mathcal{D}} - m_{KK,p} \right) \\
    &= - \frac{1}{2} \sum_{p \,\in \,\mathbb{Z}^*} \bigg(
    \log \det \left( \slashed{\mathcal{D}} - m_{KK,p} \right) + \log \det \left( - \slashed{\mathcal{D}} - m_{KK,p} \right) \bigg) \\
    &= - \frac{1}{2} \sum_{p \,\in \,\mathbb{Z}^*}
    \log \det \bigg( - \slashed{\mathcal{D}}^2 + \left[ \slashed{\mathcal{D}}, m_{KK,p} \right]
    + m_{KK,p}^2 \bigg).
\end{align}
The action is reformulated in Laplace-type form, thereby enabling the application of the heat kernel method. However, the equality in the third line is not generally valid, as it relies on the assumption that $\det A \cdot \det B = \det (AB)$, which fails for general infinite-dimensional operators. For infinite-dimensional operators, the determinant itself is ill-defined without proper regularization. Therefore, we adopt the zeta-function regularization to define the determinant:
\begin{equation}
    \begin{gathered}
    \det_{\zeta} A \coloneqq \exp (- \zeta'(s=0 \,|\, A )), \quad
    \zeta (s \,|\, A) \coloneqq
    \sum_{\{\lambda_i\} \,\in\, \text{Spec} A} \lambda_i^{-s},
    \end{gathered}
\end{equation}
with an appropriate spectral cut. Then, the failure of the multiplicative property of the determinant can be captured by the quantity
\begin{equation}
    F(A,B) \coloneqq \frac{\det_{\zeta} (AB)}{\det_{\zeta} A \cdot \det_{\zeta} B}.
\end{equation}
This function $F(A,B)$ is known as the multiplicative anomaly, and its formula is already established~\cite{Kontsevich:1994xe}. Nonetheless, this does not imply that its calculation is straightforward. In general, computing $F(A,B)$ is quite difficult, and only a few specific cases, notably for free Dirac fermion, have been calculated~\cite{Elizalde:1997nd,Elizalde:1998vd,Elizalde:1999zy,Cognola:1999xv}. When a Dirac fermion has some non-constant potential or interactions not represented by connections, one must employ perturbation theory, which further complicates the computation.


\section{Derivation of \texorpdfstring{Eq.~(\ref{eq:sfactor})}{eq:sfactor}} \label{app:fact_ol}

\subsection{The case of the KK scalar}
According to the Feynman rules, the one-loop diagram~\autoref{fig:gene_ol} is formally written as
\begin{equation}
    \mathcal{A}^{(a,b,c)}_n \sim m_{\mathrm{KK}, n}^{2(a+b)} (\phi_0)
    \int \frac{d^4p}{(2\pi)^4} \,\frac{1}{p^2+m_{\mathrm{KK}, n}^2 (\phi_0)} \frac{\big( p^2 + p \cdot k + k^2 + m_{\mathrm{KK}, n}^2 (\phi_0) \big)^c}{\prod_{I=1}^{a+b+c-1} [(p+k_I)^2 + m_{\mathrm{KK}, n}^2 (\phi_0)]}.
\end{equation}
We consider two cases (i): $a+b \leq 2$ (the momentum integral is UV divergent) and (ii): $a+b > 2$ (the momentum integral is UV convergent).

\subsubsection*{Case (i): \texorpdfstring{$a+b \leq 2$}{a few radions}}
We introduce the appropriate number of Pauli–Villars regulators~\footnote{More precisely, the regularization should be implemented by introducing additional regulator fields at the level of the Lagrangian. In the present calculation, however, we retain only the scheme-independent coefficients of one-particle Feynman diagrams, so this distinction is immaterial. A more careful treatment would be necessary when considering higher-loop effects or higher-derivative corrections.} with auxiliary UV cutoff $\Lambda_{\mathrm{aux}, n} \coloneqq n \cdot \Lambda_{\mathrm{aux}}$, if necessary. Specifically, when $a+b \leq 2$, we add $3-(a+b)$ regulators. Then, the correction becomes
\begin{align}
    \mathcal{A}^{(a,b,c)}_{n,\,a+b \leq 2} &\sim
    \Lambda_{\mathrm{aux}, n}^{2(3-(a+b))} m_{\mathrm{KK}, n}^{2(a+b)} (\phi_0) \int \frac{d^4p}{(2\pi)^4} \frac{1}{(p^2+\Lambda_{\mathrm{aux}, n}^2)^{3-(a+b)}}
    \frac{1}{p^2+m_{\mathrm{KK}, n}^2 (\phi_0)} \notag\\
    &\quad \times \prod_{I=1}^{a+b+c-1} \Delta(k_I) \times \left( p^2 + p \cdot k + k^2
    + m_{\mathrm{KK}, n}^2 (\phi_0) \right)^c,
\end{align}
where $\Delta(k_I)$ is the propagator:
\begin{equation}
    \label{eq:prop}
    \Delta(k_I) = \frac{1}{(p + k_I)^2+m_{\mathrm{KK}, n}^2 (\phi_0)}.
\end{equation}
By introducing Feynman parameters, we obtain
\begin{equation}
    \mathcal{A}^{(a,b,c)}_{n,\,a+b \leq 2} \sim \Lambda_{\mathrm{aux}, n}^{2(3-(a+b))} m_{\mathrm{KK}, n}^{2(a+b)} (\phi_0) \int dF_{c+3} \int \frac{d^4l}{(2\pi)^4} \frac{\big( l^2 + l \cdot k + k^2 + m_{\mathrm{KK}, n}^2 (\phi_0) \big)^c}{(l^2+D)^{c+3}} ,
\end{equation}
where $D=m_{\mathrm{KK}, n}^2 (\phi_0) + (x_1+\cdots + x_{3-(a+b)}) \Lambda_{\mathrm{aux}, n}^2 + k^2$. 
Moreover, we rewrite $1/(l^2+D)^{c+3}$ in terms of the proper-time parameter $u$. Then, we have
\begin{align}
    \mathcal{A}^{(a,b,c)}_{n,\,a+b \leq 2} &\sim
    \Lambda_{\mathrm{aux}, n}^{2(3-(a+b))} m_{\mathrm{KK}, n}^{2(a+b)} (\phi_0) \int dF_{c+3} \notag\\
    &\times \int_{\Lambda_{\mathrm{sp},0}^{-2}}^{\infty} du
    \int \frac{d^4l}{(2\pi)^4} \left( l^2 + l \cdot k + k^2 + m_{\mathrm{KK}, n}^2 (\phi_0) \right)^c u^{c+2} \notag\\
    &\times \exp\left[-u\left(k^2+l^2+m_{\mathrm{KK}, n}^2 (\phi_0)\right)\right] \exp\left[-u(x_1+\cdots + x_{3-(a+b)})\Lambda_{\mathrm{aux}, n}^2\right].
\end{align}
Utilizing the formula
\begin{equation}
    \int \frac{d^4l}{(2\pi)^4} \,l^{2m} e^{-ul^2} \,\sim\, u^{-2-m}
\end{equation}
and the relation $m_{\mathrm{KK}, n}^2 (\phi_0) \sim \partial_u \sim u^{-1}$, one can show that 
\begin{equation}
    \left. \left( l^2 + l \cdot k + k^2 + m_{\mathrm{KK}, n}^2 (\phi_0) \right)^c e^{-uk^2} \right|_{k^4} \,\sim\, u^{-c} k^4.
\end{equation}
Therefore, the correction reads
\begin{equation}
    \begin{split}
    \mathcal{A}^{(a,b,c)}_{n,\,a+b \leq 2} &\sim \Lambda_{\mathrm{aux}, n}^{2(3-(a+b))} m_{\mathrm{KK}, n}^{2(a+b)} (\phi_0) \int dF_{c+3} \\
    &\times \int_{\Lambda_{\mathrm{sp},0}^{-2}}^{\infty} du \,u^2 \mathcal{O}(k^4) e^{-u(x_1+\cdots + x_{3-(a+b)})\Lambda_{\mathrm{aux}, n}^2} e^{-um_{\mathrm{KK}, n}^2 (\phi_0)}.
    \end{split}
\end{equation}
Since the integration over the Feynman parameters reads
\begin{equation}
    \int dF_{c+3} \,e^{-u(x_1+\cdots + x_{3-(a+b)})\Lambda_{\mathrm{aux}, n}^2}
    \,\sim\, (u \Lambda_{\mathrm{aux}, n}^2)^{(a+b)-3},
\end{equation}
we finally obtain 
\begin{equation}
    \mathcal{A}^{(a,b,c)}_{n,\,a+b \leq 2}
    \sim m_{\mathrm{KK}, n}^{2(a+b)} (\phi_0) \mathcal{O}(k^4) \int_{\Lambda_{\mathrm{sp},0}^{-2}}^{\infty} du \,u^{a+b-1}
    e^{-um_{\mathrm{KK}, n}^2 (\phi_0)}.
\end{equation}

\subsubsection*{Case (ii): \texorpdfstring{$a+b > 2$}{many radions}}
We next consider the case of $a+b>2$, in which momentum regularization is not required. The correction is given by
\begin{align}
    \mathcal{A}^{(a,b,c)}_{n,\,a+b > 2} &\sim
    m_{\mathrm{KK}, n}^{2(a+b)} (\phi_0) \int dF_{a+b+c} \int_{\Lambda_{\mathrm{sp},0}^{-2}}^{\infty} du \int \frac{d^4l}{(2\pi)^4} \left( l^2 + l \cdot k + k^2 + m_{\mathrm{KK}, n}^2 (\phi_0) \right)^c \notag\\
    &\times u^{a+b+c-1} \exp\left[-u\left(k^2+l^2+m_{\mathrm{KK}, n}^2 (\phi_0)\right)\right],
\end{align}
where $D=m_{\mathrm{KK}, n}^2 (\phi_0) + k^2$. The previous arguments lead to the same result. 
Therefore, we deduce the factorization in Eq.~\eqref{eq:sfactor}:
\begin{equation}
    \mathcal{A}^{(a,b,c)}_n = \mathcal{A}^{(a,b,c)}_{\mathrm{single}}
    \cdot \frac{m_{\mathrm{KK}, n}^{2(a+b)} (\phi_0)}{\wGamma(a+b)} \int_{\Lambda_{\mathrm{sp},0}^{-2}}^{\infty} du \,u^{a+b-1}
    e^{-um_{\mathrm{KK}, n}^2 (\phi_0)}.
\end{equation}

\subsection{The case of the KK fermion}

We now extend the previous analysis to the case of the KK fermion, focusing in particular on periodic fermions with integer KK momentum (see \autoref{ssec:ferm_fey}). The generic one-loop Feynman diagram remains analogous to the earlier case, so that the one-loop correction can be expressed formally as
\begin{align}
    &\mathcal{A}^{(a,b,c)}_n \sim m_{\mathrm{KK}, n}^{a+b} (\phi_0)
    \int \frac{d^4p}{(2\pi)^4} \Tr \bigg[
    \frac{i \slashed{p} - m_{\mathrm{KK}, n} (\phi_0)}{p^2+m_{\mathrm{KK}, n}^2 (\phi_0)} \notag\\
    &\times \prod_{I=1}^{a+b} \Delta(k_I) \prod_{I=a+b+1}^{a+b+c-1}
    \left( \gamma \cdot (p-k) + m_{\mathrm{KK}, n} (\phi_0) \right) \Delta(k_I) \times \left( \gamma \cdot (p-k) + m_{\mathrm{KK}, n} (\phi_0) \right) \bigg],
\end{align}
where $\Delta(k_I)$ is the scalar propagator Eq.~\eqref{eq:prop}. In order to apply the proper-time method, we reformulate this as follows:
\begin{align}
    \mathcal{A}^{(a,b,c)}_n &\sim m_{\mathrm{KK}, n}^{a+b} (\phi_0)
    \int \frac{d^4p}{(2\pi)^4} \frac{1}{p^2+m_{\mathrm{KK}, n}^2 (\phi_0)} \prod_{I=1}^{a+b+c-1}
    \frac{1}{(p+k_I)^2+m_{\mathrm{KK}, n}^2 (\phi_0)} \notag\\
    &\quad \times \Tr \left[
    \left( \gamma \cdot (p-k) + m_{\mathrm{KK}, n} (\phi_0) \right)^{a+b+2c}
    \right].
\end{align}
It is important to note that the power of loop momentum in the numerator depends on whether $a+b$ is even or odd. Since our analysis is restricted to cases where $a+b \leq 4$, it is necessary to introduce $3-\lceil (a+b)/2 \rceil$ Pauli-Villars regulators, where $\lceil x \rceil$ denotes a ceiling function. Then, the correction is given by
\begin{align}
    &\mathcal{A}^{(a,b,c)}_{n, \,a+b\leq4} \sim
    \Lambda_n^{2(3-\lceil \frac{a+b}{2} \rceil)} m_{\mathrm{KK}, n}^{a+b} (\phi_0)
    \int dF_{f_{abc}} \notag\\
    &\quad \times \int \frac{d^4l}{(2\pi)^4} \Tr \left[
    \left( \gamma \cdot (p-k) + m_{\mathrm{KK}, n} (\phi_0) \right)^{a+b+2c}
    \right] \int_{\Lambda_{\mathrm{sp},0}^{-2}}^{\infty} du \,u^{f_{abc}-1} e^{-u(l^2+D)}, 
\end{align}
where 
\begin{gather}
    f_{abc} \coloneqq 3-\lceil \frac{a+b}{2} \rceil +a+b+c, \\
    D = m_{\mathrm{KK}, n}^2 (\phi_0) + (x_1+\cdots + x_{3-\lceil \frac{a+b}{2} \rceil}) \Lambda_n^2 + k^2.
\end{gather}
The trace part, leading to the four-derivative corrections, reads
\begin{align}
    &\Tr \left.\left[ \left( \gamma \cdot (p-k) + m_{\mathrm{KK}, n} (\phi_0) \right)^{a+b+2c}
    \right]\right|_{\leq \,k^4} \notag\\
    &= \sum_{s=0}^{c+\lfloor \frac{a+b}{2} \rfloor}
    (\gamma \cdot l)^{2s} m_{\mathrm{KK}, n}^{a+b+2c-2s} (\phi_0) + \sum_{s=1}^{c+\lfloor \frac{a+b}{2} \rfloor} (\gamma \cdot l)^{2(s-1)} (\gamma \cdot k)^2 m_{\mathrm{KK}, n}^{a+b+2c-2s} (\phi_0) \notag\\
    &+ \sum_{s=2}^{c+\lfloor \frac{a+b}{2} \rfloor} (\gamma \cdot l)^{2(s-2)} (\gamma \cdot k)^4 m_{\mathrm{KK}, n}^{a+b+2c-2s} (\phi_0),
\end{align}
where $\lfloor x \rfloor$ is the floor function. Combining this with $\exp(-u(k^2+l^2))$ and performing the $l$-integration, we find
\begin{align}
    \Tr \left[ \left( \gamma \cdot (p-k) + m_{\mathrm{KK}, n} (\phi_0) \right)^{a+b+2c} \right]
    e^{-uk^2} &\sim\, \sum_s u^{-s} k^4 m_{\mathrm{KK}, n}^{a+b+2c-2s} (\phi_0) \\
    &\sim\, u^{-c} k^4 m_{\mathrm{KK}, n}^{a+b} (\phi_0), 
\end{align}
where, in the final line, we have used the relation $m_{\mathrm{KK}, n}^{2} (\phi_0) \sim u^{-1}$. Applying the same arguments in the scalar case, we can derive Eq.~\eqref{eq:sfactor}:
\begin{equation}
    \label{eq:ferm_fact}
    \mathcal{A}^{(a,b,c)}_{n, \,a+b \leq 4}
    = \mathcal{A}^{(a,b,c)}_{\mathrm{single}}
    \cdot \frac{m_{\mathrm{KK}, n}^{2(a+b)} (\phi_0)}{\wGamma(a+b)} \int_{\Lambda_{\mathrm{sp},0}^{-2}}^{\infty} du \,u^{a+b-1}
    e^{-um_{\mathrm{KK}, n}^2 (\phi_0)}.
\end{equation}
This result is identical to the scalar result.


\section{Details of one-loop computation}\label{app:one-loop}

In this section, we summarize the four-derivative matter-loop corrections for the minimal scalar and Dirac tower models defined in Appendix~\ref{app:cc}, including operators involving the KK photon. We first present the results, and then briefly outline the derivation procedure for the most technically involved diagrams. Other Feynman diagrams can be evaluated more straightforwardly following the same procedure. The Supplemental Material~\cite{SupplementalMaterial} contains the intermediate expressions and the results of the momentum integrals used in these computations.

\subsection{KK scalar}

The four-dimensional minimal scalar tower action is
\begin{equation}
    S = \frac{M_{\mathrm{pl}, 4}^{2}}{2} \int d^4 x \sqrt{-g} \,\bigg[ \mathcal{R} - (\nabla \phi)^2 + \frac{1}{4 g^2(\phi)} F^{\mu \nu} F_{\mu \nu} - \sum_{n \in \,\mathbb{Z}} \left( | D \varphi_n |^2 + m_{\mathrm{KK}, n}^2 (\phi) |\varphi_n|^2 \right) \bigg],
\end{equation}
where
\begin{equation}
    g (\phi) = m_{KK} (\phi), \quad D_{\mu} \varphi_n = (\partial_{\mu} - i n A_{\mu}) \varphi_n,
\end{equation}
and the associated field strength is given by
\begin{equation}
    \Omega_{\mu \nu} = [D_{\mu}, D_{\nu}] = - i n F_{\mu \nu}.
\end{equation}
To determine the coefficient of $\mathcal{R}^2$, we first estimate the relevant order $2k$ of higher-dimensional correction. A generic term can be formally expressed as
\begin{equation}
    \mathcal{R}^{\,a} \,\nabla^{\,2b} \,\Omega^{\,2c} \,m_{\mathrm{KK}}^{\,2d}
\end{equation}
subject to the constraints on the number of derivatives and mass dimensions:
\begin{align}
    4 &= 2 (a+b+c),\label{eq:nd_b}\\
    2k &= 2 (a+b+2c+d), \label{eq:md_b}
\end{align}
respectively. Since the KK mass $m_{\mathrm{KK}}$ always appears under differentiation, the parameter $b$ must satisfy $b \geq [ (d+1)/2 ]$. Combining this with Eq.~\eqref{eq:nd_b} yields the inequality $d \leq 4 - 2c$. Consequently, the constraint Eq.~\eqref{eq:md_b} implies that it is sufficient to consider $k \leq 6$.

\begin{figure}[t]
    \centering

    \newlength{\diagramheight}
    \setlength{\diagramheight}{0.31\textheight}

    \begin{minipage}[t][\diagramheight][t]{0.48\textwidth}
        \centering

        \includegraphics[width=0.24\linewidth]{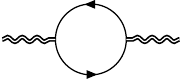}

        \smallskip
        $(k=2)$

        \medskip

        \begin{minipage}[t]{0.23\linewidth}
            \centering
            \includegraphics[width=\linewidth]{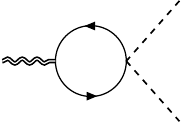}
        \end{minipage}\hfill
        \begin{minipage}[t]{0.23\linewidth}
            \centering
            \includegraphics[width=\linewidth]{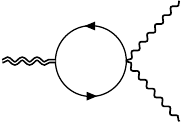}
        \end{minipage}\hfill
        \begin{minipage}[t]{0.23\linewidth}
            \centering
            \includegraphics[width=\linewidth]{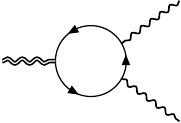}
        \end{minipage}\hfill
        \begin{minipage}[t]{0.23\linewidth}
            \centering
            \raisebox{0.5em}{\includegraphics[width=\linewidth]{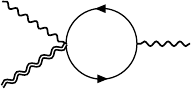}}
        \end{minipage}

        \smallskip
        $(k=3)$

        \medskip

        \begin{minipage}[t]{0.23\linewidth}
            \centering
            \includegraphics[width=\linewidth]{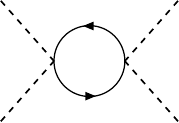}
        \end{minipage}\hfill
        \begin{minipage}[t]{0.23\linewidth}
            \centering
            \includegraphics[width=\linewidth]{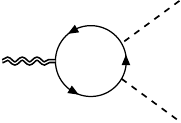}
        \end{minipage}\hfill
        \begin{minipage}[t]{0.23\linewidth}
            \centering
            \raisebox{0.5em}{\includegraphics[width=\linewidth]{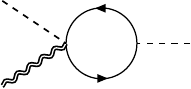}}
        \end{minipage}\hfill
        \begin{minipage}[t]{0.23\linewidth}
            \centering
            \includegraphics[width=\linewidth]{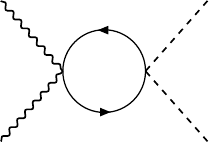}
        \end{minipage}

        \smallskip

        \begin{minipage}[t]{0.23\linewidth}
            \centering
            \raisebox{-0.8em}{\includegraphics[width=\linewidth]{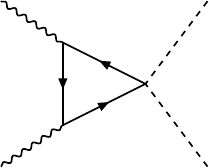}}
        \end{minipage}\hfill
        \begin{minipage}[t]{0.23\linewidth}
            \centering
            \raisebox{-0.7em}{\includegraphics[width=\linewidth]{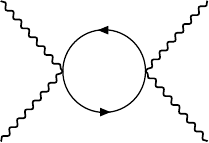}}
        \end{minipage}\hfill
        \begin{minipage}[t]{0.23\linewidth}
            \centering
            \raisebox{-1em}{\includegraphics[width=\linewidth]{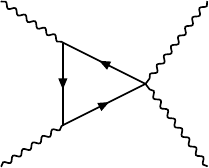}}
        \end{minipage}\hfill
        \begin{minipage}[t]{0.23\linewidth}
            \centering
            \raisebox{-1.2em}{\includegraphics[width=\linewidth]{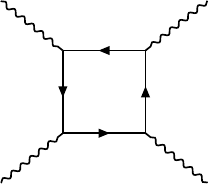}}
        \end{minipage}

        \smallskip
        $(k=4)$
    \end{minipage}
    \hfill
    \begin{minipage}[t][\diagramheight][c]{0.48\textwidth}
    \vspace*{-3em}
        \centering

        \begin{minipage}[t]{0.88\linewidth}
            \centering
            \begin{minipage}[t]{0.23\linewidth}
                \centering
                \raisebox{0.25em}{\includegraphics[width=\linewidth]{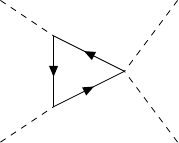}}
            \end{minipage}\hfill
            \begin{minipage}[t]{0.23\linewidth}
                \centering
                \raisebox{0.25em}{\includegraphics[width=\linewidth]{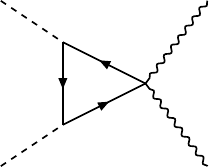}}
            \end{minipage}\hfill
            \begin{minipage}[t]{0.23\linewidth}
                \centering
                \includegraphics[width=\linewidth]{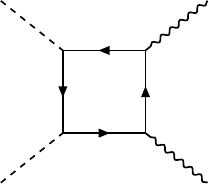}
            \end{minipage}\hfill
            \begin{minipage}[t]{0.23\linewidth}
                \centering
                \includegraphics[width=\linewidth]{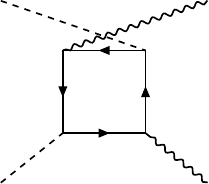}
            \end{minipage}

            \smallskip
            $(k=5)$
        \end{minipage}

        \medskip

        \begin{minipage}[t]{0.28\linewidth}
            \centering
            \includegraphics[width=0.85\linewidth]{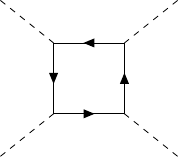}

            \smallskip
            $(k=6)$
        \end{minipage}
    \end{minipage}

    \caption{One-loop diagrams contributing at $k=2,\ldots,6$.}

    \par\smallskip
    \noindent\textbf{Alt text:}
    A collection of 18 one-loop Feynman diagrams grouped
by $k$: one diagram for $k=2$, four for $k=3$,
eight for $k=4$, four for $k=5$, and one for $k=6$.
Solid lines with arrows represent KK scalars forming
circular, triangular, or square loops. Dashed lines
represent radions, and double-wavy lines represent
gravitons. The diagrams show different combinations
and arrangements of radion and graviton interactions
with the scalar loops.
    \label{fig:k2tok6}
\end{figure}

Let us first derive the Feynman rules in Euclidean signature%
\footnote{To avoid unnecessary complications arising from the imaginary unit $i$, the calculation is performed in the Euclidean signature. For a discussion of Euclidean spinors and Wick-rotation conventions, see~\cite{vanNieuwenhuizen:1996tv}. Here, we briefly comment on the correspondence with the results in the Lorentzian signature. The Wick rotation is applicable only to the matter sector in a flat background. Consequently, in a general background, one cannot use Euclidean calculations to directly derive Lorentzian signature results through an inverse Wick rotation. 

However, in certain cases, equivalent results in both signatures can be obtained without performing a Wick rotation. This occurs in the one-loop calculation as follows. Initially, relative factors for propagators and vertices are $I_L/I_E=-i$ and $V_L/V_E=i$, respectively. Since the number of propagators and the number of vertices are equal in a one-loop diagram, the relative factor for the actions is $i\Gamma_L/(-\Gamma_E) = i \cdot (-i)^{\#I} i^{\#V} = i$.  Therefore, the Lorentzian action can be obtained from the Euclidean action by simply reversing its overall sign.}%
. The Euclidean action is given by
\begin{equation}
    S_E = \frac{M_{\mathrm{pl}, 4}^2}{2} \int d^4x \,\sqrt{g} \bigg[ \mathcal{R} + (\nabla \phi)^2 - \frac{1}{4g^2(\phi)} F_{\mu \nu} F^{\mu \nu} + \sum_{n \in \,\mathbb{Z}} \left( | D \varphi_n |^2 + m_{\mathrm{KK}, n}^2 (\phi) |\varphi_n|^2 \right) \bigg].
\end{equation}
Then, the matter part of the action for $n \geq 1$ can be expanded as
\begin{align}
    S_E &= M_{\mathrm{pl}, 4}^2 \sum_{n \geq 1} \int \bigg[ \varphi^*_n (- \partial^2 + m_{KK,n}^2(\phi_0)) \,\varphi_n \notag\\
    &+ \big(\partial_{\phi_0} m_{KK,n}^2(\phi_0) \,\phi
    + \frac{1}{2} \partial_{\phi_0}^2 m_{KK,n}^2(\phi_0) \,\phi^2 \big)
    \,|\varphi_n|^2 \notag\\
    &- inA_{\mu} (\varphi_n \partial^{\mu} \varphi_n^* - \varphi_n^* \partial^{\mu} \varphi_n) + n^2 A_{\mu} A^{\mu} |\varphi_n|^2 \notag\\
    &- \oh^{\mu \nu} \bigg(\partial_{\mu} \varphi^*_n \partial_{\nu} \varphi_n + in \left( A_{\mu} \varphi_n^* \partial_{\nu} \varphi_n
    - A_{\nu} \varphi_n \partial_{\mu} \varphi_n^* \right) \notag\\
    &+ \frac{1}{2} \delta_{\mu \nu}
    \left( m_{KK,n}^2(\phi_0)
    + \partial_{\phi_0} m_{KK,n}^2(\phi_0) \,\phi \right) |\varphi_n|^2 \bigg) + \cdots  \bigg], 
\end{align}
where $\oh_{\mu \nu}$ is the trace-reversed metric defined by $\oh_{\mu \nu} = h_{\mu \nu} - 1/2 \,\delta_{\mu \nu} h$. The Feynman vertices are defined as follows.
\begin{align}
    \label{eq:FR_pp}
    \raisebox{-.88\height}{\includegraphics[width=3em]{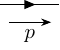}} \quad &: \quad \frac{1}{p^2+m_{\mathrm{KK}, n}^2 (\phi_0)}, \\
    \raisebox{-.88\height}{\includegraphics[width=3em]{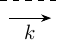}} \quad &: \quad \frac{1}{k^2}, \\
    \parbox{3.5em}{\includegraphics[width=3.5em]{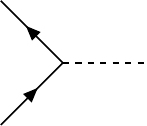}} \quad &: \quad - \sqrt{6} \,m_{\mathrm{KK}, n}^2 (\phi_0), \\
    \parbox{3em}{\includegraphics[width=3em]{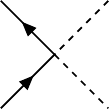}} \quad &:\quad - \sqrt{6}^2 m_{\mathrm{KK}, n}^2 (\phi_0), \\
    \label{eq:FR_gKK}
    \parbox{4.5em}{\includegraphics[width=4.5em]{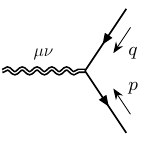}} \quad &: \quad \frac{1}{2} (-p_{\mu} q_{\nu} - p_{\nu} q_{\mu} + \delta_{\mu \nu} m_{\mathrm{KK}, n}^2(\phi_0)), \\
    \label{eq:FR_gpKK}
    \parbox{4em}{\includegraphics[width=4em]{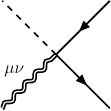}} \quad &: \quad \frac{\sqrt{6}}{2} \delta_{\mu \nu} m_{\mathrm{KK}, n}^2 (\phi_0). \\
    \label{eq:FR_AKK}
    \parbox{4.5em}{\includegraphics[width=4.5em]{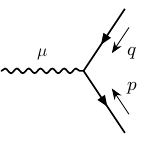}} \quad &: \quad n (q_{\mu} - p_{\mu}), \\
    \parbox{4em}{\includegraphics[width=4em]{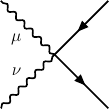}} \quad &: \quad -2n^2 \delta_{\mu \nu}, \\
    \label{eq:FR_AgKK}
    \parbox{4.5em}{\includegraphics[width=4.5em]{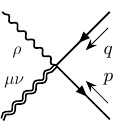}} \quad &: \quad -n ( \delta_{\mu \rho} q_{\nu} - \delta_{\nu \rho} p_{\mu})
\end{align}
The relevant diagrams are shown in~\autoref{fig:k2tok6}.
The one-loop corrections at each $k$ in the Lorentzian signature are given by
\begin{align}
    S_{\text{1-loop}}^{\,k=2} &= \frac{1}{(4 \pi)^2} \int d^4 x \sqrt{-g} \left( \sqrt{\pi} N(\phi_0) - \log N(\phi_0) \right) \notag\\
    &\quad \times \left( \frac{1}{2! \cdot 6^2} \mathcal{R}^2 - \frac{1}{180} \mathcal{R}^{\mu \nu} \mathcal{R}_{\mu \nu} + \frac{1}{180} \mathcal{R}_{\mu \nu \rho \sigma} \mathcal{R}^{\mu \nu \rho \sigma} \right), \\
    \label{eq:s_k=3}
    S_{\text{1-loop}}^{\,k=3} &= \frac{1}{(4 \pi)^2} \int d^4 x \sqrt{-g} \,\sqrt{\pi} N(\phi_0) \bigg[  \frac{-1}{5} \mathcal{R} (\nabla \phi)^2 \notag\\
    &\quad + \frac{1}{g^2(\phi_0)} \bigg( \frac{1}{60} F\indices{_{\mu \nu}^{; \nu}} F\indices{^{\mu \rho}_{; \rho}} - \frac{1}{180} \mathcal{R}_{\mu \nu \rho \sigma} F^{\mu \nu} F^{\rho \sigma} - \frac{1}{90} \mathcal{R}_{\mu \nu} F^{\mu \rho} F\indices{^{\nu}_{\rho}} - \frac{1}{72} \mathcal{R} F_{\mu \nu} F^{\mu \nu} \bigg) \bigg], \\
    \label{eq:s_k=4}
    S_{\text{1-loop}}^{\,k=4} &= \frac{1}{(4 \pi)^2} \int d^4 x \sqrt{-g} \,\sqrt{\pi} N(\phi_0) \bigg[ \frac{3}{10} (\nabla \phi)^4 + \frac{7}{60} \mathcal{R} (\nabla \phi)^2 - \frac{1}{30} \mathcal{R}^{\mu \nu} \phi_{; \mu} \phi_{; \nu} \notag\\
    &\quad + \frac{1}{g^2(\phi_0)} \frac{7}{120} F_{\mu \nu} F^{\mu \nu} (\nabla \phi)^2 + \frac{1}{g^4(\phi_0)} \left( \frac{1}{288} \left( F_{\mu \nu} F^{\mu \nu} \right)^2 + \frac{1}{360} F_{\mu \rho} F\indices{_{\nu}^{\rho}} F^{\mu \sigma} F\indices{^{\nu}_{\sigma}} \right) \bigg], \\
    \label{eq:s_k=5}
    S_{\text{1-loop}}^{\,k=5} &= \frac{1}{(4 \pi)^2} \int d^4 x \sqrt{-g} \,\sqrt{\pi} N(\phi_0) \bigg[ \frac{-2}{5} (\nabla \phi)^4 \notag\\
    &\quad + \frac{1}{g^2(\phi_0)} \Big( \frac{-1}{30} F_{\mu \nu} F^{\mu \nu} (\nabla \phi)^2 + \frac{1}{15} F_{\mu \rho} F\indices{_{\nu}^{\rho}} \phi^{;\mu} \phi^{;\nu} \Big) \bigg], \\
    \label{eq:s_k=6}
    S_{\text{1-loop}}^{\,k=6} &= \frac{1}{(4 \pi)^2}
    \int d^4 x \sqrt{-g} \,\sqrt{\pi} N(\phi_0) \frac{3}{20} (\nabla \phi)^4. 
\end{align}
The scalar-loop terms involving only the graviton and photon have been calculated in~\cite{Cheung:2014ega,Dunne:2004nc}. Combining the terms gives the scalar-tower contribution at four-derivative order:
\begin{align}
    S_{\text{1-loop}}^{\,\partial^4} &= \frac{1}{(4 \pi)^2} \int d^4 x \sqrt{-g} \bigg\{\sqrt{\pi} N(\phi_0) \bigg[ \frac{1}{120} \mathcal{R}^2 + \frac{1}{60} \mathcal{R}^{\mu \nu} \mathcal{R}_{\mu \nu}
    - \frac{1}{12} \mathcal{R} (\nabla \phi)^2 \notag\\
    &\quad - \frac{1}{30} \mathcal{R}^{\mu \nu} \phi_{; \mu} \phi_{; \nu} + \frac{1}{20} (\nabla \phi)^4 - \frac{1}{180} \mathcal{R}_{\mu \nu \rho \sigma} F^{\mu \nu} F^{\rho \sigma} - \frac{1}{90} \mathcal{R}_{\mu \nu} F^{\mu \rho} F\indices{^{\nu}_{\rho}} \notag\\
    &\quad - \frac{1}{72} \mathcal{R} F_{\mu \nu} F^{\mu \nu} + \frac{1}{40} F_{\mu \nu} F^{\mu \nu} (\nabla \phi)^2 + \frac{1}{15} F_{\mu \rho} F\indices{_{\nu}^{\rho}} \phi^{;\mu} \phi^{;\nu} + \frac{1}{288} \left( F_{\mu \nu} F^{\mu \nu} \right)^2 \notag\\
    &\quad + \frac{1}{360} F_{\mu \rho} F\indices{_{\nu}^{\rho}} F^{\mu \sigma} F\indices{^{\nu}_{\sigma}} \bigg] - \log N(\phi_0) \left[ \frac{1}{120} \mathcal{R}^2 + \frac{1}{60} \mathcal{R}^{\mu \nu} \mathcal{R}_{\mu \nu} \right] \bigg\},
\end{align}
where we have ignored the terms with undifferentiated radion. We also absorb the gauge coupling into the field strength. Since the charged source is absent, $F\indices{_{\mu \nu}^{; \nu}} F\indices{^{\mu \rho}_{; \rho}}$ vanishes. 


\subsection{KK fermion}
\label{ssec:ferm_fey}

We now consider the four-dimensional minimally coupled Dirac tower specified in Eq.~\eqref{eq:minimal_tower_action}. Its Euclidean action at $A_\mu=0$ is given in Eq.~\eqref{eq:ferm_ac}; the photon is restored through $\mathcal{D}_\mu=\nabla_\mu-ipA_\mu$, with no Pauli vertex.

We expand the matter action around the flat metric $g_{\mu \nu} = \delta_{\mu \nu} + h^{GW}_{\mu \nu}$ and derive the expansion of the matter part of action~\cite{Godazgar:2018boc}. If we set the perturbation of the vielbein as $e_{\mu}^a = \delta_{\mu}^a + 1/2 \, h_{\mu}^a$, one finds that $h^{GW}_{\mu \nu}=h\indices{_{(\mu}^a} e\indices{_{\nu) a}} \eqqcolon h_{(\mu \nu)}$. The only non-trivial part is a perturbation of the spin connection $\omega\indices{_{\mu}^{ab}}$. We can show that this does not contribute to the result as follows. Focusing only on the symmetric part of $h_{\mu \nu}$, one finds $\delta \omega\indices{_{\mu}^{ab}} = -2 \partial\indices{^{[a}} h\indices{_{\mu}^{b]}}$. Therefore, we have
\begin{align}
    \delta S &\supset \frac{M_{\mathrm{pl}, 4}^2}{2} \int d^4x \,e \left( \opsi \gamma^{\mu}  \frac{1}{4} \delta \omega\indices{_{\mu}^{ab}} \gamma_{ab} \psi + \text{h.c.} \right) \\
    &= \frac{M_{\mathrm{pl}, 4}^2}{2} \int d^4x \,e \frac{1}{4} \opsi \left\{ \gamma^{\mu}, \gamma^{ab} \right\} \psi \,(-2 \partial_{a} h_{b \mu}),
\end{align}
and this vanishes. This reminds us of the 1.5 order formalism in supergravity. 

Before deriving the Feynman rules, we consider the sum over $p \in \mathbb{Z}^* = (\mathbb{Z}+s/2) \setminus \{ 0 \}$ appearing in the one-loop effective action, which can formally be written as
\begin{equation}
    S_{\text{1-loop}} = \sum_{p \,\in \,\mathbb{Z}^*} \int d^4 x \sqrt{g} \,\sum_{a,b,c} \mathcal{A}^{(a,b,c)}_{p, \,a+b \leq 4} \,\phi^{a+2b} \,h_{\mu \nu}^c.
\end{equation}
This comes down to examining the following infinite sum:
\begin{equation}
    \sum_{p \,\in \,\mathbb{Z}^*} p^{2m} e^{-p^2/N^2(\phi_0)}, \quad m \geq 0.
\end{equation}
For $m=0$, if a zero mode is present, its contribution cannot be neglected. Consequently, the subleading behavior depends on the value of $s$. In particular, for anti-periodic boundary conditions, there is no zero mode, and hence no $\log N(\phi_0)$ term appears. By contrast, the leading behavior can be extracted from the theta function
\begin{equation}
    \vartheta_{s,0} (q) = \sum_{p \,\in \,\mathbb{Z}+s/2} e^{-p^2/N^2(\phi_0)}.
\end{equation}
by successive differentiation with respect to $N^{-2}(\phi_0)$. 
By applying the Poisson resummation, one finds that the leading term is the same for $s=0$ and $s=1$. In fact,
\begin{align}
    \vartheta_{s,0} (q) &= \sqrt{\pi}N(\phi_0)
    \,\vartheta_{0, s} (q'= e^{- \pi^2N^2(\phi_0)}) = \sqrt{\pi}N(\phi_0)
    \,\sum_{n \,\in \,\mathbb{Z}} (-1)^{sn} e^{- \pi^2 n^2N^2(\phi_0)} \\
    &= \sqrt{\pi}N(\phi_0) + \mathcal{O} (e^{- \pi^2N^2(\phi_0)}).
\end{align}
In this paper, we therefore focus on the case $s=0$, for which the $\log N(\phi_0)$ term is present. 

The matter part of the action can be expanded as
\begin{align}
    &S_E = M_{\mathrm{pl}, 4}^2 \sum_{n \,\in\, \mathbb{Z}^*} \int
    \bigg[ \frac{1}{2} \opsi_n \slasholr{D} \psi_n
    - m_{\mathrm{KK}, n} (\phi_0) \,\opsi_n \psi_n \notag\\
    &\quad - \bigg(\partial_{\phi_0} m_{\mathrm{KK}, n} (\phi_0) \,\phi
    + \frac{1}{2} \partial_{\phi_0}^2 m_{\mathrm{KK}, n} (\phi_0) \,\phi^2
    + \mathcal{O}(\phi^3)\bigg) \,\opsi_n \psi_n \notag\\
    &\quad - \oh^{\mu \nu} \bigg( \frac{1}{4} \opsi \gamma_{(\mu} \olr{D}_{\nu)} \psi + \delta_{\mu \nu} \bigg( \frac{1}{8} \opsi \,\slasholr{D}\, \psi - \frac{1}{2} \big(m_{\mathrm{KK}, n} (\phi_0)
    + \partial_{\phi_0} m_{\mathrm{KK}, n} (\phi_0) \phi\big) \,\opsi_n \psi_n \bigg) \bigg) \bigg].
\end{align}
The relevant propagators and vertices displayed below follow from this minimal action:
\begin{align}
    \raisebox{-.88\height}{\includegraphics[width=3em]{Figures/prop_KK.pdf}} \quad &: \quad \frac{i \slashed{p} - m_{\mathrm{KK}, n} (\phi_0)}{p^2+m_{KK,n}^2 (\phi_0)}, \\
    \raisebox{-.88\height}{\includegraphics[width=3em]{Figures/prop_radion.pdf}} \quad &: \quad \frac{1}{k^2}, \\
    \parbox{4em}{\includegraphics[width=4em]{Figures/3pt_vertex.pdf}} \quad &: \quad \frac{\sqrt{6}}{2} m_{\mathrm{KK}, n} (\phi_0), \\
    \parbox{3.5em}{\includegraphics[width=3.5em]{Figures/4pt_vertex.pdf}} \quad &:\quad \left( \frac{\sqrt{6}}{2} \right)^2 m_{\mathrm{KK}, n} (\phi_0), \\
    \parbox{5em}{\includegraphics[width=5em]{Figures/gff_3pt.pdf}} \quad &: \quad \frac{i}{4} \gamma_{(\mu} (p-q)_{\nu)} + \delta_{\mu \nu} \left( \frac{i}{8} (\slashed{p} - \slashed{q}) - \frac{1}{2} m_{\mathrm{KK}, n} (\phi_0) \right), \\
    \parbox{4em}{\includegraphics[width=4em]{Figures/gffs_4pt.pdf}} \quad &: \quad - \frac{\sqrt{6}}{4} \delta_{\mu \nu} m_{\mathrm{KK}, n} (\phi_0), \\
    \parbox{4.5em}{\includegraphics[width=4.5em]{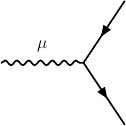}} \quad &: \quad i n \,\gamma_{\mu}, \\
    \parbox{4em}{\includegraphics[width=4em]{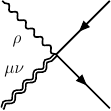}} \quad &: \quad -in \,\frac{3}{4} \delta_{(\mu \nu} \gamma_{\rho)}.
\end{align}

The relevant Feynman diagrams are those obtained from the diagrams appearing in the case of the KK scalar, with the exclusion of those containing an $A^2$-vertex. The one-loop corrections at each $k$ in the Lorentzian signature, apart from the minimally coupled fermion-loop terms involving only the graviton and photon~\cite{Cheung:2014ega,Dunne:2004nc}, are given by
\begin{align}
    S_{\text{1-loop}}^{\,k=2} &= \frac{1}{(4 \pi)^2} \int d^4 x \sqrt{-g} \,\bigl(\sqrt{\pi} N(\phi_0) - \log N(\phi_0) \bigr)
    \left( \frac{-1}{30} \mathcal{R}^2
    + \frac{1}{10} \mathcal{R}^{\mu \nu} \mathcal{R}_{\mu \nu} \right), \label{eq:f_k=2}\\
    S_{\text{1-loop}}^{\,k=3} &= \frac{1}{(4 \pi)^2} \int d^4 x \sqrt{-g} \,\sqrt{\pi} N(\phi_0) \frac{-1}{10} \mathcal{R} (\nabla \phi )^2, \label{eq:f_k=3}\\
    S_{\text{1-loop}}^{\,k=4} &= \frac{1}{(4 \pi)^2} \int d^4 x \sqrt{-g} \,\sqrt{\pi} N(\phi_0) \bigg( \frac{9}{20} (\nabla \phi)^4 + \frac{11}{60} \mathcal{R} (\nabla \phi)^2 - \frac{11}{30} \mathcal{R}_{\mu \nu} \phi^{; \mu} \phi^{; \nu} \notag\\
    &\quad + \frac{1}{g^2(\phi_0)} \frac{2}{15} F_{\mu \nu} F^{\mu \nu} (\nabla \phi)^2 \bigg), \label{eq:f_k=4}\\
    S_{\text{1-loop}}^{\,k=5} &= \frac{1}{(4 \pi)^2} \int d^4 x \sqrt{-g} \,\sqrt{\pi} N(\phi_0) \bigg[  \frac{-11}{20} (\nabla \phi)^4 \notag\\
    &\quad + \frac{1}{g^2(\phi_0)} \bigg( \frac{11}{15} F_{\mu \nu} F^{\mu \nu} (\nabla \phi)^2 - \frac{7}{30} F_{\mu \rho} F\indices{_{\nu}^{\rho}} \phi^{;\mu} \phi^{;\nu} \bigg) \bigg], \label{eq:f_k=5}\\
    S_{\text{1-loop}}^{\,k=6} &= \frac{1}{(4 \pi)^2} \int d^4 x \sqrt{-g} \,\sqrt{\pi} N(\phi_0) \frac{11}{40} (\nabla \phi)^4. \label{eq:f_k=6}
\end{align}
Combining the terms gives the minimally coupled fermion-tower contribution at four-derivative order:
\begin{align}
    S_{\text{1-loop}}^{\,\partial^4} &= \frac{1}{(4 \pi)^2} \int d^4 x \sqrt{-g} \bigg\{ \sqrt{\pi} N(\phi_0) \bigg[  - \frac{1}{30} \mathcal{R}^2 + \frac{1}{10} \mathcal{R}^{\mu \nu} \mathcal{R}_{\mu \nu} + \frac{1}{12} \mathcal{R} (\nabla \phi)^2 \notag\\
    &\quad - \frac{11}{30} \mathcal{R}^{\mu \nu} \phi_{; \mu} \phi_{; \nu} + \frac{7}{40} (\nabla \phi)^4 + \frac{1}{45} \mathcal{R}_{\mu \nu \rho \sigma} F^{\mu \nu} F^{\rho \sigma} - \frac{13}{45} \mathcal{R}_{\mu \nu} F^{\mu \rho} F\indices{^{\nu}_{\rho}} \notag\\
    &\quad + \frac{1}{18} \mathcal{R} F_{\mu \nu} F^{\mu \nu} + \frac{13}{15} F_{\mu \nu} F^{\mu \nu} (\nabla \phi)^2 - \frac{7}{30} F_{\mu \rho} F\indices{_{\nu}^{\rho}} \phi^{;\mu} \phi^{;\nu} + \frac{1}{45} \left( F_{\mu \nu} F^{\mu \nu} \right)^2 \notag\\
    &\quad + \frac{7}{180} \left(F_{\mu \nu} \widetilde{F}^{\mu \nu} \right)^2 \bigg] - \log N(\phi_0) \left[ - \frac{1}{30} \mathcal{R}^2 + \frac{1}{10} \mathcal{R}^{\mu \nu} \mathcal{R}_{\mu \nu} \right] \bigg\},
\end{align}
where we have ignored the terms with undifferentiated radion. We also absorb the gauge coupling into the field strength. Since the charged source is absent, $F\indices{_{\mu \nu}^{; \nu}} F\indices{^{\mu \rho}_{; \rho}}$ vanishes. 

\subsection{The photon-radion correction for the KK scalar at \texorpdfstring{$k=5$}{10-mass dimensions}}

In this subsection, we present the derivation of the photon–radion corrections for the case of $k=5$ in the KK scalar sector, which represents the most technically involved example among the Feynman diagram computations. 

The relevant diagrams are 
\begin{center}
\includegraphics[width=0.25\linewidth]{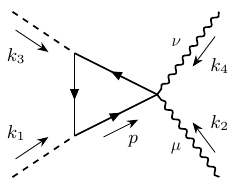}
\hfill
\includegraphics[width=0.24\linewidth]{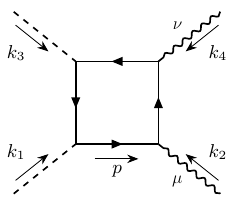}
\hfill
\includegraphics[width=0.31\linewidth]{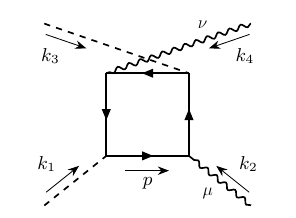}.
\end{center}
For simplicity of formulas, we use the following notation
\begin{equation}
    \begin{gathered}
    a = k_1^2, \quad b = k_1 \cdot (k_2+k_4), \quad t = -(k_2+k_4)^2, \\
    d = k_2^2, \quad f = k_{12}, \quad g = k_2 \cdot (k_2+k_4).
    \end{gathered}
\end{equation}
The contribution of the left diagram is given by
\begin{align}
    &- \mathcal{A}_{\mu, \nu}^{\,(\phi, \phi, A^2)} = 2 \int \frac{d^4 p}{(2 \pi)^4}
    \frac{1}{p^2+m_{\mathrm{KK}, n}^2 (\phi_0)} \notag\\
    &\quad \times \frac{1}{(p+k_2+k_4)^2+m_{\mathrm{KK}, n}^2 (\phi_0)}
    \frac{1}{(p-k_1)^2+m_{\mathrm{KK}, n}^2 (\phi_0)} \times (-\sqrt{6} m_{\mathrm{KK}, n}^2 (\phi_0))^2 (-2n^2 \delta_{\mu \nu}) \notag\\
    &= - \frac{1}{(4\pi)^2 m_{\mathrm{KK}}^2(\phi_0)}
    \frac{2}{15} \delta_{\mu \nu}  (3a^2+6ab-5at+4b^2-6bt +3t^2).
\end{align}
The factor 2 in the first line accounts for the complex degrees of freedom of the KK scalar. The contribution of the middle diagram is given by
\begin{align}
    - \mathcal{A}_{\mu, \nu}^{\,(\phi, \phi, A, A)} &= 2 \times 2 \int \frac{d^4 p}{(2 \pi)^4}
    \frac{1}{p^2+m_{\mathrm{KK}, n}^2 (\phi_0)}
    \frac{1}{(p+k_2)^2+m_{\mathrm{KK}, n}^2 (\phi_0)} \notag\\
    &\quad \times \frac{1}{(p+k_2+k_4)^2+m_{\mathrm{KK}, n}^2 (\phi_0)} \frac{1}{(p-k_1)^2+m_{\mathrm{KK}, n}^2 (\phi_0)} \notag\\
    &\quad \times (-\sqrt{6} m_{\mathrm{KK}, n}^2 (\phi_0))^2 \,n(2p+k_2)_{\mu} \,n(2p+2k_2+k_4)_{\nu} \notag\\
    &= 12 n^2 m_{\mathrm{KK}, n}^2 (\phi_0)^2 \int dF_4 \int \frac{d^4 l}{(2 \pi)^4} \frac{1}{(l^2+D)^4} (2p+k_2)_{\mu} (2p+2k_2+k_4)_{\nu}, 
\end{align}
where 
\begin{align}
    l &= p +x_2 k_2 + x_3 (k_2+k_4) - x_4 k_1, \\
    D &= (x2-x2^2) d + (x_3-x_3^2) (-t) + (x_4 - x_4^2) a - 2 x_2 x_3 g + 2x_2 x_4 f + 2x_3 x_4 b + m_{\mathrm{KK}, n}^2 (\phi_0).
\end{align}
The additional factor of 2 comes from the fact that, in two of the three possible configurations, adjacent vertices are of the same type.
One gets
\begin{align}
    &- \mathcal{A}_{\mu, \nu}^{\,(\phi, \phi, A, A)} = \frac{1}{(4\pi)^2m_{\mathrm{KK}}^2(\phi_0)} \bigg[ \delta_{\mu \nu} \frac{4}{105} \Big( 6 a^2 +4 b^2+8 d f-8 dg-9 d t+6 d^2-4 f g-5 f t \notag\\
   &+4f^2+8 g t+4 g^2+6 t^2 +a (8 b+9 d+8 f-5 g-9 t) +b (5 d+4 f-4 g-8 t) \Big) \notag\\
   &+ k_{1\mu} k_{1\nu} \frac{8}{105} ( -9 a-6 b-5 d-6 f+2 g+5 t ) \notag\\
   &+ k_{1 \mu} k_{2\nu} \frac{4}{105} ( -26 a-16 b-15 d-16 f+6 g+15 t ) \notag\\
   &+ (k_2+k_4)_{\mu} k_{1 \nu} \frac{4}{105} ( 8 a+8 b+5 d+4 f-4 g-8 t ) \notag\\
   &+ (k_2+k_4)_{\mu} k_{2 \nu} \frac{4}{105} ( 13 a+12 b+10 d+6 f-8 g-16 t ) \bigg].
\end{align}
The contribution of the right diagram is given by
\begin{align}
    - \mathcal{A}_{\mu, \nu}^{\,(\phi, \phi, A, A)^C} &= 2 \int \frac{d^4 p}{(2 \pi)^4}
    \frac{1}{p^2+m_{\mathrm{KK}, n}^2 (\phi_0)}
    \frac{1}{(p+k_2)^2+m_{\mathrm{KK}, n}^2 (\phi_0)} \notag\\
    &\quad \times \frac{1}{(p+k_2+k_3)^2+m_{\mathrm{KK}, n}^2 (\phi_0)} \frac{1}{(p-k_1)^2+m_{\mathrm{KK}, n}^2 (\phi_0)} \notag \\
    &\quad \times (-\sqrt{6} m_{\mathrm{KK}, n}^2 (\phi_0))^2 \,n(2p+k_2)_{\mu} \,n(2p+k_2+k_3-k_1)_{\nu} \notag\\
    &= 12 n^2 m_{\mathrm{KK}, n}^2 (\phi_0)^2 \int dF_4 \int \frac{d^4 l}{(2 \pi)^4} \frac{1}{(l^2+D)^4} (2p+k_2)_{\mu} (2p-2k_1-k_4)_{\nu}, 
\end{align}
where 
\begin{align}
    l &= p +x_2 k_2 - x_3 (k_1+k_4) - x_4 k_1, \\
    D &= (x2-x2^2) d + (x_3-x_3^2) (a+d-2f-t-2g+2b) \notag\\
    &+ (x_4 - x_4^2) a - 2 x_2 x_3 (d-f-g) + 2x_2 x_4 f + 2x_3 x_4 (f-a-b) + m_{\mathrm{KK}, n}^2 (\phi_0).
\end{align}
One gets
\begin{align}
    &- \mathcal{A}_{\mu, \nu}^{\,(\phi, \phi, A, A)^C} = \frac{1}{(4\pi)^2m_{\mathrm{KK}}^2(\phi_0)} \bigg[ \delta_{\mu \nu} \frac{2}{105} \Big( 9 a^2 +12 b^2-18 d g-13 d t+9 d^2+6 f g+3 f t \notag\\
   &+6 f^2+16 g t+12 g^2+6 t^2 + a (18 b+17 d-17 g-13 t) +b (17 d-6 f-20 g-16 t) \Big) \notag\\
   &+ k_{1\mu} k_{1\nu} \frac{4}{105} ( 18 a+18 b+17 d-17 g-13 t ) \notag\\
   &+ k_{1 \mu} k_{2\nu} \frac{2}{105} ( 18 a+18 b+17 d+6 f-14 g-13 t ) \notag\\
   &+ (k_2+k_4)_{\mu} k_{1 \nu} \frac{2}{105} (18 a+24 b+17 d-6 f-20 g-16 t) \notag\\
   &+ (k_2+k_4)_{\mu} k_{2 \nu} \frac{2}{105} ( 9 a+12 b+8 d-8 g-8 t ) \bigg].
\end{align}
Therefore, we have
\begin{align}
    &- \mathcal{A}_{\mu, \nu}^{\,(\phi, \phi, A^2)} - \mathcal{A}_{\mu, \nu}^{\,(\phi, \phi, A, A)} - \mathcal{A}_{\mu, \nu}^{\,(\phi, \phi, A, A)^C} \notag\\
    &= \frac{1}{(4\pi)^2m_{\mathrm{KK}}^2(\phi_0)} \bigg[ \delta_{\mu \nu} \frac{2}{105} \Big( a (-8 b+35 d+16 f-27 g +4 t) \notag\\
    &+b (27 d+2f-28 g+10 t) -8 b^2+16 d f-34 d g \notag\\
    &-31d t+21 d^2-2 f g-7 f t+14 f^2+32g t+20 g^2-3 t^2 \Big) \notag\\
    &+ k_{1\mu} k_{1\nu} \frac{4}{105} ( 6 b+7 d-12 f-13 g-3 t ) \notag\\
    &+ k_{1 \mu} (k_1+k_3)_{\nu} \frac{2}{105} ( 34 a+14 b+13 d+26 f+2 g-17 t ) \notag\\
    &+ (k_1+k_3)_{\mu} k_{1 \nu} \frac{2}{105} ( -34 a-40 b-27 d-2 f+28 g+32 t ) \notag\\
    &+ (k_1+k_3)_{\mu} (k_1+k_3)_{\nu} \frac{2}{105} \Big( 35 a+4 (9 b+7 d+3 f-6 g-10 t) \Big) \bigg].
\end{align}
The generic formula for the one-loop corrections is
\begin{align}
    - \mathcal{L}_{gen} &= \frac{\sqrt{g}}{(4 \pi)^2 m_{\mathrm{KK}}^2(\phi_0)}
    \Big( c_1 F_{\mu \nu} F^{\mu \nu} (\nabla \phi)^2 + c_2 F_{\mu \nu} F^{\mu \nu} \phi \Delta \phi + c_3 F_{\mu \rho} F\indices{_{\nu}^{\rho}} \phi^{;\mu} \phi^{;\nu} \notag\\
    &\quad + c_4 F_{\mu \rho} F\indices{_{\nu}^{\rho}} \phi^{;\mu \nu} \phi + c_5 F\indices{^{\mu \nu}_{;\nu}} F\indices{_{\mu \rho}^{;\rho}} \phi \phi \Big).
\end{align}
Defining the momentum expression as
\begin{equation}
    F_{\mu \nu} F^{\mu \nu} (\nabla \phi)^2 = \frac{1}{4} \left[ F_{\mu \nu} F^{\mu \nu} (\nabla \phi)^2 \right]_k \phi A^{\mu} \phi A^{\nu},
\end{equation}
those candidates can be expressed in the momentum language as
\begin{align}
    &\left[ F_{\mu \nu} F^{\mu \nu} (\nabla \phi)^2 \right]_k = 8 \Big((a+b)(d-g) \delta_{\mu \nu} + (a+b) (k_1+k_3)_{\mu} (k_1+k_3)_{\nu} \Big),\\
    &\left[ F_{\mu \nu} F^{\mu \nu} \phi \Delta \phi \right]_k = 8 \Big(a(g-d) \delta_{\mu \nu} - a (k_1+k_3)_{\mu} (k_1+k_3)_{\nu} \Big),\\
    &\left[ F_{\mu \rho} F\indices{_{\nu}^{\rho}} \phi^{;\mu} \phi^{;\nu} \right]_k = 4 \Big( f(t+f+g-b) \delta_{\mu \nu} +(d-g) k_{1\mu} k_{1\nu} \notag\\
    &+(t+f+2g-b-d) k_{1 \mu} (k_1+k_3)_{\nu} - f (k_1+k_3)_{\mu} k_{1 \nu} + f (k_1+k_3)_{\mu} (k_1+k_3)_{\nu} \Big), \\
    &\left[ F_{\mu \rho} F\indices{_{\nu}^{\rho}} \phi^{;\mu \nu} \phi \right]_k \notag \\
    &= 4 \Big( f(b-f) \delta_{\mu \nu} + (g-d) k_{1\mu} k_{1\nu} + (b-f) k_{1 \mu} (k_1+k_3)_{\nu} + f (k_1+k_3)_{\mu} k_{1 \nu} \Big),\\
    &\left[ F\indices{^{\mu \nu}_{;\nu}} F\indices{_{\mu \rho}^{;\rho}} \phi \phi \right]_k = 4 \delta_{\mu \nu} d(d-t-2g). 
\end{align}
These momentum-space expressions have many redundancies associated with the exchange $k_2 \leftrightarrow k_4$ or $k_1 \leftrightarrow k_3$. Accompanied by $\delta_{\mu \rho}$, we have the following relations
\begin{gather}
    dg-g^2 = \frac{t}{2} (g-d), \quad (t-2b)I=0, \quad 
    b^2 = \frac{t}{2}(a+b) - ab, \\
    (f, g) J = \left(\frac{b}{2}, \frac{-t}{2} \right) J, \quad
    f(d-g) = \frac{b}{4} (t+2d), \quad
    dg = -2fg-\frac{1}{2} t(b+d)
\end{gather}
where $I$ is a variable invariant under $k_1 \leftrightarrow k_3$, and $J$ is a variable invariant under $k_2 \leftrightarrow k_4$. Focusing on the terms with $\delta_{\mu \nu}$, one finds
\begin{equation}
    c_1 = \frac{1}{60}, \quad c_2 = - \frac{1}{15}, \quad c_3 = - \frac{1}{5},
    c_4 = - \frac{4}{15}, \quad c_5 = \frac{1}{10}.
\end{equation}
This indeed cancels the terms with $k_{1\mu} k_{1\nu}$. For $k_{1 \mu} (k_1+k_3)_{\nu}$ and $(k_1+k_3)_{\mu} k_{1 \nu}$, one can show that
\begin{equation}
    k_{1 \mu} (k_1+k_3)_{\nu} U(k_1,k_2,k_3,k_4) = (k_1+k_3)_{\mu} k_{1 \nu} U(k_1,k_4,k_3,k_2),
\end{equation}
where $U$ represents a generic variable. Applying the following relation
\begin{equation}
    k_{1 \mu} (k_1+k_3)_{\nu} (2g+t) = 0,
\end{equation}
we obtain
\begin{align}
    &(4\pi)^2m_{\mathrm{KK}}^2(\phi_0) \bigg(- \mathcal{A}_{\mu, \nu}^{\,(\phi, \phi, A^2)} - \mathcal{A}_{\mu, \nu}^{\,(\phi, \phi, A, A)} - \mathcal{A}_{\mu, \nu}^{\,(\phi, \phi, A, A)^C} - (- \mathcal{L}_{gen}) \bigg)\notag\\
    &= k_{1 \mu} (k_1+k_3)_{\nu} \frac{-16}{15} d + (k_1+k_3)_{\mu} (k_1+k_3)_{\nu}
    \frac{2}{105} ( 29b + 28d + 54f - 24g - 40t ) \\
    &= k_{2\nu} k_{4\mu} \frac{2}{105} ( 29b + 54f - 24g - 40t ),
\end{align}
which vanishes. Thus, the $k=5$ photon-radion correction is given by
\begin{align}
    S_{\text{1-loop}}^{\,k=5} &= - \frac{\sqrt{\pi} N(\phi_0)}{(4 \pi)^2 g^2(\phi_0)}
    \int d^4 x \sqrt{g} \,\Big( \frac{1}{60} F_{\mu \nu} F^{\mu \nu} (\nabla \phi)^2 - \frac{1}{15} F_{\mu \nu} F^{\mu \nu} \phi \Delta \phi - \frac{1}{5} F_{\mu \rho} F\indices{_{\nu}^{\rho}} \phi^{;\mu} \phi^{;\nu} \notag\\
    &\quad - \frac{4}{15} F_{\mu \rho} F\indices{_{\nu}^{\rho}} \phi^{;\mu \nu} \phi + \frac{1}{10} F\indices{^{\mu \nu}_{;\nu}} F\indices{_{\mu \rho}^{;\rho}} \phi \phi \Big). \\
    &= - \frac{\sqrt{\pi} N(\phi_0)}{(4 \pi)^2 g^2(\phi_0)}
    \int d^4 x \sqrt{g} \,\Big( - \frac{1}{30} F_{\mu \nu} F^{\mu \nu} (\nabla \phi)^2 + \frac{1}{15} F_{\mu \rho} F\indices{_{\nu}^{\rho}} \phi^{;\mu} \phi^{;\nu} \Big).
\end{align}


\section{Species scale for a \texorpdfstring{$d$}{d}-dimensional string tower} \label{app:string_tower}

We now consider the one-loop correction from a string tower. Unlike the KK tower, string excitations include higher-spin states and exhibit an exponentially growing degeneracy at each level. These features make it difficult to compute the exact heat-kernel coefficients for the full string spectrum. We therefore restrict ourselves to the leading parametric behavior in a scalar approximation with the step-function proper-time cutoff used in the main text. To proceed, we make two simplifying assumptions.

First, we approximate all string excitations by scalar fields. This approximation is not meant to reproduce the exact heat-kernel coefficients of the full string spectrum, since higher-spin fields have different curvature couplings and, in supersymmetric spectra, nontrivial cancellations may occur among bosonic and fermionic states. Nevertheless, for the purpose of estimating the leading scaling of the one-loop correction, we use the scalar result as a proxy and assume that spin-dependent effects do not alter the parametric dependence on the string level.

Second, we use the asymptotic formula for the degeneracy of the $d$-dimensional string excitations,
\begin{equation}
    d_n \sim n^{-\frac{d+1}{2}} e^{c \sqrt{n}},
\end{equation}
where $c$ is an $\mathcal{O}(1)$ constant depending on the specific type of string theory, and an overall constant in $d_n$ is suppressed. This formula is valid only for sufficiently large $n$. As we will see below, highly excited states dominate the regulated sum in the limit considered here, so the precise degeneracy at lower levels does not affect its leading exponential behavior.

With these assumptions, we can write the one-loop correction as
\begin{align}
    S_{\text{1-loop}} &= \int d^d x \sqrt{-g} \,\sum_{n \geq 1} d_n \int_{\Lambda_{\mathrm{sp}, 0}^{-2}}^{\infty} \frac{du}{u} \,\text{tr}_V e^{-u (- \nabla^2 + m_n^2 (\phi_0 + \phi))} \\
    &= \int d^d x \sqrt{-g} \,\sum_{k, j \geq 0} \widetilde{e}_{2k}^{\,(j)}(x; \nabla, \phi) \sum_{n \geq 1} d_n m_n^{2j}(\phi_0) \int_{\Lambda_{\mathrm{sp}, 0}^{-2}}^{\infty} du \,u^{k-d/2-1} e^{-u m_n^2 (\phi_0)}.
    \label{eq:one_loop_string_smaller}
\end{align}
Here, $m_n^2 (\phi) = n M_s^2 (\phi)$ is the mass squared of the $n$-th string excitation, where $\phi$ is the $d$-dimensional dilaton. An overall numerical factor in the mass spectrum is absorbed into $M_s$. We introduce the provisional loop cutoff $\Lambda_{\mathrm{sp}, 0}$ and examine the limit $M_{s,0}/\Lambda_{\mathrm{sp},0}\to0$. The scale inferred below is a conditional estimate within this regulated scalar model. For simplicity, we only consider even $d>2$.

For $k \geq d/2+1$, the one-loop formula can be expressed as
\begin{align}
    S_{\text{1-loop}}^{\,k} &\simeq \int d^d x \sqrt{-g} \,\sum_{j \geq 0} \widetilde{e}_{2k}^{\,(j)} (x; \nabla, \phi) \Gamma(k-\frac{d}{2}) M_{s, 0}^{d-2k+2j} \notag\\
    & \qquad \times \sum_{l=0}^{k-\frac{d}{2}-1} \frac{t^l}{l!} \sum_{n \geq 1} n^{- \frac{d+1}{2}+j+l-k+d/2} \,e^{- n t + c \sqrt{n}},
\end{align}
where $t \coloneqq (M_{s, 0}/\Lambda_{\mathrm{sp}, 0})^2$, and the asymptotic degeneracy has been substituted. Following the approach in Eq.~\eqref{eq:ol_high}, we obtain
\begin{align}
    -\partial_t \sum_{l=0}^{k-\frac{d}{2}-1} \frac{t^l}{l!} \sum_{n \geq 1} n^{- \frac{d+1}{2}+j+l-k+d/2} \,e^{- n t + c \sqrt{n}}
    = \frac{t^{k-\frac{d}{2}-1}}{(k-\frac{d}{2}-1)!} \sum_{n \geq 1} n^{- \frac{d+1}{2}+j} \,e^{- n t + c \sqrt{n}}.
\end{align}
The infinite sum on the right-hand side behaves as
\begin{equation}
    \label{eq:infinite_sum_string}
    \sum_{n \geq 1} n^{- \frac{d+1}{2}+j} \,e^{- n t + c \sqrt{n}}
    \simeq t^{d-2j-\frac{1}{2}} \,e^{c^2 / (4t)}
\end{equation}
in the emergent string limit $t\to0^+$ for fixed $d$, $j$, and $c$. Here, we suppress the constant prefactor and subleading corrections. The derivation, including the prefactor, is given in Appendix~\ref{app:infinite_sum_string}.

The exponential factor has its maximum at
\begin{equation}
    n_*=\frac{c^2}{4t^2},
    \qquad
    \frac{m_{n_*}(\phi_0)}{\Lambda_{\mathrm{sp},0}}
    =\frac{c}{2\sqrt{t}}\longrightarrow\infty.
\end{equation}
The power of $n$ in the summand shifts the saddle only by a relative $\mathcal{O}(t)$ correction. Thus, although the dominant levels are indeed highly excited, their masses lie parametrically above the provisional cutoff. The exponential degeneracy compensates for the proper-time suppression. The resulting tower contribution is therefore a weighted sum over the spectrum, rather than a count of particles with masses below $\Lambda_{\mathrm{sp},0}$.

To obtain the action, one may use the large-argument expansion
\begin{equation}
    \Gamma(q,z)=z^{q-1}e^{-z}\bigl[1+\mathcal{O}(z^{-1})\bigr],
    \qquad q=k-\frac{d}{2},
\end{equation}
which is valid for any fixed $q$ as $z\to+\infty$. The relevant saddle satisfies $z=nt\sim t^{-1}\gg1$. Applying Appendix~\ref{app:infinite_sum_string} to the resulting sum gives
\begin{align}
    S_{\text{1-loop}}^{\,k} \simeq \int d^d x \sqrt{-g} \,\sum_{j \geq 0}
    C_{d,j}\,\widetilde{e}_{2k}^{\,(j)} (x; \nabla, \phi)
    \left(\frac{\Lambda_{\mathrm{sp}, 0}}{M_{s, 0}}\right)^{2j-2d-1}
    \Lambda_{\mathrm{sp}, 0}^{2j}
    \frac{\exp{\left(\frac{c^2}{4} \left( \frac{\Lambda_{\mathrm{sp}, 0}}{M_{s, 0}} \right)^{2}\right)}}{\Lambda_{\mathrm{sp}, 0}^{2k-d}},
\end{align}
where
\begin{equation}
    C_{d,j}=c\sqrt{\pi}\left(\frac{2}{c}\right)^{d+3-2j}
\end{equation}
in the normalization of the asymptotic degeneracy written above. This result holds at each fixed $k$ and $j$, including $k\leq d/2$. In particular, no singular factor $\Gamma(k-d/2)$ is needed in the latter cases. For $k>d/2$, the explicit $\Gamma(k-d/2)$ in the finite-sum representation cancels the factorial in its dominant term.

In the limit $M_{s,0}/\Lambda_{\mathrm{sp},0}\to0$, the common exponential enhancement dominates the displayed algebraic factors. To compare with the KK estimate, we now retain only this exponential factor and introduce the weight
\begin{equation}
    N \coloneqq \exp{\left(\frac{c^2}{4} \left( \frac{\Lambda_{\mathrm{sp}, 0}}{M_{s, 0}} \right)^{2}\right)}.
\end{equation}
Here $N$ denotes the weight in this exponential-only approximation, not the cumulative number of string states below the cutoff. Assuming that the remaining matching contributions do not parametrically dominate or cancel the tower term, we impose the same parametric perturbative relation as in the KK case. With order-one factors omitted, the resulting reduced equation is
\begin{equation}
    N = \exp{\left( \frac{c^2}{4} \left( \frac{M_{\mathrm{pl}, d}}{M_{s, 0}} \right)^{2} N^{\frac{-2}{d-2}} \right)}.
\end{equation}
The solution of this reduced equation is
\begin{equation}
    N^{\frac{2}{d-2}} = \exp{\left( W_0 \left( \frac{c^2}{4} \frac{2}{d-2} \left( \frac{M_{\mathrm{pl}, d}}{M_{s, 0}} \right)^{2} \right) \right)},
\end{equation}
where $W_0 (x)$ is the principal branch of the Lambert $W$ function. Thus, within this approximation,
\begin{equation}
    N \simeq \frac{1}{(-2 \log g_{s,d})^{\frac{d-2}{2}}} \cdot \frac{1}{g_{s,d}^2} \,\lesssim \,\frac{1}{g_{s,d}^2},
\end{equation}
where order-one numerical factors are suppressed and
\begin{equation}
    \frac{1}{g_{s,d}^2} \coloneqq \left( \frac{M_{\mathrm{pl}, d}}{M_{s, 0}} \right)^{d-2}.
\end{equation}
The corresponding candidate scale is
\begin{align}
    \Lambda_{\mathrm{sp}, 0} &= M_{\mathrm{pl}, d} \exp{\left( - \frac{1}{2}  W_0 \left(\frac{c^2}{4} \frac{2}{d-2} \left( \frac{M_{\mathrm{pl}, d}}{M_{s, 0}} \right)^{2} \right) \right)} \\
    &\simeq (\log N)^{\frac{1}{2}} M_{s, 0}.
    \label{eq:species_scale_string}
\end{align}
The Lambert-$W_0$ expressions solve the reduced equation, rather than the full threshold sum. Restoring the algebraic factors changes subleading logarithmic terms in the inferred scale. For fixed operator indices, these factors shift $(\Lambda_{\mathrm{sp},0}/M_{s,0})^2$ by terms of order $\log\log(M_{\mathrm{pl},d}/M_{s,0})$, while leaving its leading growth proportional to $\log(M_{\mathrm{pl},d}/M_{s,0})$ unchanged.

Taking $M_{s,0}$ to have the asymptotic form $M_{\mathrm{pl},d}\exp(-\phi_0/\sqrt{d-2})$, with a $\phi_0$-independent prefactor, the slope in this reduced estimate behaves as
\begin{align}
    \left| \frac{\partial_{\phi_0} \Lambda_{\mathrm{sp}, 0}}{\Lambda_{\mathrm{sp}, 0}} \right| &= \left| \frac{1}{2} \partial_{\phi_0} W_0 \left(\frac{c^2}{4} \frac{2}{d-2} \left( \frac{M_{\mathrm{pl}, d}}{M_{s, 0}} \right)^{2} \right) \right| \notag\\
    &= \frac{1}{\sqrt{d-2}} \left| \frac{\log N^{\frac{2}{d-2}}}{1 + \log N^{\frac{2}{d-2}}} \right| \notag\\
    &= \frac{1}{\sqrt{d-2}} - \frac{1}{2} \cdot \frac{1}{\phi_0} + o \left(\frac{1}{\phi_0} \right) \notag\\
    &<\, \frac{1}{\sqrt{d-2}},
\end{align}
satisfying the bound~\cite{vandeHeisteeg:2023ubh, vandeHeisteeg:2023dlw} at least asymptotically.

As discussed in~\cite{Castellano:2022bvr}, if the maximal excitation level is denoted by $N_s$, a naive counting of the string states below the species scale gives
\begin{equation}
    \sum_{n = 1}^{N_s} e^{\sqrt{n}} .
\end{equation}
Defining the number of species from this counting leads to
\begin{equation}
    \Lambda_{\mathrm{sp}} \simeq \log N \cdot M_{s} .
\end{equation}
This differs from the expression in Eq.~\eqref{eq:species_scale_string} by a factor of $(\log N)^{1/2}$.  The logarithm here is associated with a sharp mass cutoff, whereas the square-root logarithm in Eq.~\eqref{eq:species_scale_string} comes from the proper-time weighted sum, dominated by masses above its provisional cutoff. The difference reflects these distinct treatments of the exponentially growing spectrum.

The square-root logarithmic enhancement in Eq.~\eqref{eq:species_scale_string} is parametrically consistent with estimates of the high-energy scale at which the string tree-level approximation ceases to dominate~\cite{Mende:1989wt,Basile:2023blg}. This agreement motivates a comparison with string amplitudes, but a physical identification requires the full string spectrum, its interactions, and the appropriate ultraviolet completion of the regulated sum.


\section{Derivation of the formula~\texorpdfstring{Eq.~(\ref{eq:infinite_sum_string})}{(infinite sum formula)}} \label{app:infinite_sum_string}

We will show the asymptotic formula
\begin{equation} \label{eq:infinite_sum_string_appendix}
    \sum_{n \geq 1} n^{-l} e^{-t n + c \sqrt{n}} \,\simeq \,c \sqrt{\pi} \left( \frac{2}{c} \right)^{2l} t^{2l-\frac{3}{2}} \, e^{c^2/(4t)}, \quad l \in \frac{\mathbb{Z}}{2},
\end{equation}
as $t\to0^+$, with $c>0$ and $l$ fixed. We use the Euler--Maclaurin formula at fixed order and bound its remainder relative to the leading term. The argument treats integer and half-integer $l$ in the same way.

For $f_l(x)=x^{-l}e^{-tx+c\sqrt{x}}$, the Euler--Maclaurin formula to order $2m$ reads
\begin{align} \label{eq:EM_m}
    \sum_{n \geq 1}^N f_l(n)
    &=\int_1^N f_l(x)\,dx+\frac{f_l(1)+f_l(N)}{2}
    \notag\\
    &\quad+\sum_{k=1}^m\frac{B_{2k}}{(2k)!}
    \left( \partial^{2k-1} f_l(N) - \partial^{2k-1} f_l (1) \right)+R_{2m},
\end{align}
where $B_{2k}$ are the Bernoulli numbers. The remainder is
\begin{equation}
    R_{2m}=-\int_1^N b_{2m}(x) \partial^{2m} f_l(x)\,dx,
\end{equation}
where $b_{2m} (x) = B_{2m} (x-[x]) / (2m)!$. Here, $B_{2m}(x)$ is the Bernoulli polynomial. With the inequality~\cite{knopp1990theory}
\begin{equation}
    | b_{2m} (x) | \leq \frac{4}{(2 \pi)^{2m}},
\end{equation}
we have
\begin{equation}
    | R_{2m} | \leq \frac{4}{(2 \pi)^{2m}} \int_1^N | \partial^{2m} f_l(x) | dx.
\end{equation}
At each fixed $t>0$, all derivatives of $f_l$ vanish at infinity. We may therefore take $N\to\infty$ while keeping $m$ fixed.

We first evaluate the integral
\begin{align}
    I_l(t)&\coloneqq\int_1^\infty x^{-l}e^{-tx+c\sqrt{x}}\,dx
    \notag\\
    &=2e^{c^2/(4t)}\int_1^\infty
    y^{1-2l}\exp\!\left[-t\left(y-\frac{c}{2t}\right)^2\right]dy,
\end{align}
where $x=y^2$. The saddle is at $y=c/(2t)$, with width of order $t^{-1/2}$. Expanding the power of $y$ about this saddle and performing the Gaussian integral yields
\begin{equation}
    I_l(t)=c\sqrt{\pi}\left(\frac{2}{c}\right)^{2l}
    t^{2l-\frac{3}{2}}e^{c^2/(4t)}
    \bigl[1+\mathcal{O}(t)\bigr].
\end{equation}
The term linear in the displacement integrates to zero at leading order; the first relative correction is of order $t$. Contributions from the lower endpoint are exponentially small relative to the saddle contribution.

It remains to control the Euler--Maclaurin remainder. The first nontrivial order, $m=1$, is sufficient:
\begin{equation}
    \sum_{n\geq1}f_l(n)
    =I_l(t)+\frac{f_l(1)}{2}-\frac{f_l'(1)}{12}+R_2,
    \qquad
    |R_2|\leq\frac{1}{12}\int_1^\infty|f_l''(x)|\,dx.
\end{equation}
To bound this integral, define
\begin{equation} 
    \psi_l(x)=\log f_l(x),\qquad
    \psi_l'(x)=-t+\frac{c}{2\sqrt{x}}-\frac{l}{x},
    \qquad
    \psi_l''(x)=-\frac{c}{4x^{3/2}}+\frac{l}{x^2}.
\end{equation}
Using $f_l''=[(\psi_l')^2+\psi_l'']f_l$ and integrating the first term by parts, we obtain
\begin{align}
    \int_1^\infty|f_l''(x)|\,dx
    &\leq |f_l'(1)|+2\int_1^\infty|\psi_l''(x)|f_l(x)\,dx
    \notag\\
    &\leq |f_l'(1)|+\frac{c}{2}I_{l+3/2}(t)
    +2|l|I_{l+2}(t).
\end{align}
The integral asymptotics give $I_{l+3/2}/I_l=\mathcal{O}(t^3)$ and $I_{l+2}/I_l=\mathcal{O}(t^4)$, while $f_l(1)$ and $f_l'(1)$ remain bounded as $t\to0^+$. Consequently,
\begin{equation}
    \sum_{n\geq1}f_l(n)=I_l(t)\bigl[1+\mathcal{O}(t^3)\bigr],
\end{equation}
which proves Eq.~\eqref{eq:infinite_sum_string_appendix}. The required statement is that the remainder is small relative to the leading term; its absolute value need not vanish. No limit $m\to\infty$ is required.

For $l=0$, the integral can also be evaluated explicitly. The finite-order Euler--Maclaurin formula then takes the exact form
\begin{align}
    \sum_{n\geq1}e^{-tn+c\sqrt{n}}
    &=\left[\frac{1}{t}+\frac{1}{2}
    -\frac{1}{12}\left(\frac{c}{2}-t\right)\right]e^{c-t} +\frac{c\sqrt{\pi}}{2t^{3/2}}e^{c^2/(4t)}
    \left[1+\operatorname{erf}\!\left(\frac{c-2t}{2\sqrt{t}}\right)\right]
    +R_2
    \notag\\
    &\simeq\frac{c\sqrt{\pi}}{t^{3/2}}e^{c^2/(4t)}.
\end{align}
Here $R_2$ is the remainder defined above with $l=0$. This representation retains the closed form of the integral while keeping the discrete-sum correction explicit.

\end{document}